\documentclass[aps,prb,twocolumn,floatfix,superscriptaddress,longbibliography]{revtex4-1}
\usepackage{amsmath,graphicx,txfonts,bm}
\usepackage[colorlinks,citecolor=magenta,linkcolor=blue]{hyperref}

\begin{document}

\title{Barycentric rational function approximation made simple: A fast analytic continuation method for Matsubara Green's functions}

\author{Li Huang}
\email{huangli@caep.cn}
\affiliation{Science and Technology on Surface Physics and Chemistry Laboratory, P.O. Box 9-35, Jiangyou 621908, China}

\author{Changming Yue}
\affiliation{Department of Physics, Southern University of Science and Technology, Shenzhen 518055, China}
\affiliation{Guangdong Provincial Key Laboratory of Advanced Thermoelectric Materials and Device Physics, Southern University of Science and Technology, Shenzhen, 518055, China}

\date{\today}

\begin{abstract}
Analytic continuation is a critical step in quantum many-body computations, connecting imaginary-time or Matsubara Green's functions with real-frequency spectral functions, which can be directly compared to experimental results. However, due to the ill-posed nature of the analytic continuation problems, they have not been completely solved so far. In this paper, we suggest a simple, yet highly efficient method for analytic continuations of Matsubara Green's functions. This method takes advantage of barycentric rational functions to directly interpolate Matsubara Green's functions. At first, the nodes and weights of the barycentric rational functions are determined by the adaptive Antoulas-Anderson algorithm, avoiding reliance on the non-convex optimization. Next, the retarded Green's functions and the relatively spectral functions are evaluated by the resulting interpolants. We systematically explore the performance of this method through a series of toy models and realistic examples, comparing its accuracy and efficiency with other popular methods, such as the maximum entropy method. The benchmark results demonstrate that the new method can accurately reproduce not only continuous but also discrete spectral functions, irrespective of their positive definiteness. It works well even in the presence of intermediate noise, and outperforms traditional analytic continuation methods in computational speed. We believe that this method should stand out for its robustness against noise, broad applicability, high precision, and ultra efficiency, offering a promising alternative to the maximum entropy method.
\end{abstract}

\maketitle

%%\tableofcontents

\section{Introduction\label{sec:intro}}

At finite-temperature quantum many-body calculations, the outputs of quantum Monte Carlo methods~\cite{gubernatis_kawashima_werner_2016,RevModPhys.83.349,RevModPhys.73.33}, many-body perturbation theory~\cite{Aryasetiawan_1998,PhysRev.139.A796,RevModPhys.74.601,PhysRevB.91.235114}, and lattice gauge theory~\cite{PhysRevD.24.2278,PhysRevB.24.4295,ASAKAWA2001459} are often imaginary-time Green's functions $G(\tau)$ or Matsubara Green's functions $G(i\omega_n)$. They are not directly linked to observable quantities. Thus, if we intend to compare them with experimental data, we must convert them to the real axis to obtain the retarded Green's functions $G(\omega)$ and then extract the corresponding spectral functions $A(\omega)$. This procedure is known as analytic continuation. It is evident that analytic continuation provides a bridge between quantum many-body calculations and experimental observables~\cite{many_body_book,many_body_book_2016}.

Mathematically speaking, $A(\omega)$ is related to $G(\tau)$ or $G(i\omega_n)$ through the following Laplace transformation~\cite{many_body_book}:
\begin{equation}
\label{eq:fredholm}
G(x) = \int dy~K(x,y) A(y),
\end{equation}
where $K(x,y)$ is the so-called kernel function or kernel matrix (let $x = \tau$ or $i\omega_n$, $y = \omega$). Given $A(y)$, one can easily derive the matching $G(x)$ by numerical integration. Nevertheless, analytic continuation is a typical inverse problem~\cite{PhysRevB.34.4744,PhysRevLett.55.1204}. Its input is $G(x)$. One has to solve the above integral equation (sometimes it is called the Fredholm integral equation in the literature) to get the solution $A(y)$. It is not a trivial task. The bottleneck lies in the fact that $A(y)$ is very sensitive to $G(x)$. On one hand, a tiny perturbation or fluctuation in $G(x)$ can often lead to a significant change in the corresponding $A(y)$. On the other hand, $G(\tau)$ or $G(i\omega_n)$ obtained from finite-temperature quantum Monte Carlo simulations usually contain sizable random noise~\cite{gubernatis_kawashima_werner_2016,RevModPhys.83.349,RevModPhys.73.33}. To make matters worse, some data points in $G(\tau)$ or $G(i\omega_n)$ occasionally might be filtered out due to the autocorrelation effect. These factors pose severe challenges to analytic continuation calculations~\cite{PhysRevB.41.2380}.

Over the past few decades, people have developed quite a few analytic continuation methods, which can be roughly divided into two categories. (1) \emph{Fitting-based methods}. These methods try to parameterize the spectral functions by a large number of $\delta$-like functions or a set of orthogonal basis (such as singular vectors of the kernel matrix $K$), and then fit the Green's functions. Such that analytic continuation problems are transformed into numerical optimization problems, which can be solved by using different stochastic or non-stochastic algorithms~\cite{optimization_book}. Typical fitting-based methods include the maximum entropy method (MaxEnt)~\cite{PhysRevB.41.2380,JARRELL1996133,Bryan1990,PhysRevB.44.6011,PhysRevB.81.155107,PhysRevE.94.023303,PhysRevB.98.205102,PhysRevB.96.155128,PhysRevB.95.121104,PhysRevB.92.060509}, stochastic analytic continuation (SAC)~\cite{beach2004,PhysRevB.57.10287,PhysRevE.94.063308,PhysRevB.76.035115,PhysRevX.7.041072,SHAO20231}, etc. Note that the MaxEnt method is the workhorse of the field, since it realizes a good balance between efficiency and accuracy. The SAC method has evolved many variants~\cite{PhysRevB.102.035114,PhysRevB.101.085111,PhysRevB.78.174429}, such as the stochastic analytic inference (SAI)~\cite{PhysRevE.81.056701}, stochastic optimization method (SOM)~\cite{PhysRevB.62.6317,PhysRevB.95.014102,KRIVENKO2019166,KRIVENKO2022108491}, stochastic pole expansion (SPX)~\cite{PhysRevB.108.235143,PhysRevD.109.054508}, and so on. These SAC-like methods are quite popular because they are stable and not sensitive to data noise. But they are very time-consuming. In order to resolve the subtle features in the spectra, tremendous resources should be allocated. (2) \emph{Interpolation-based methods}. These methods aim to interpolate, rather than fit, Matsubara Green's functions using some sorts of rational functions. Once the analytic form of the Matsubara Green's function in the whole complex plane is established [i.e., $G(z)$ with $z \in \mathbb{C}$], one can obtain the retarded Green's function $G(\omega)$ and the corresponding spectral function $A(\omega)$ through a simple variable substitution ($z \to \omega + i\eta$). Typical methods in this class include the Pad\'{e} approximation (PA)~\cite{PhysRevB.87.245135,PhysRevB.93.075104,Vidberg1977,PhysRevD.96.036002,PhysRevB.61.5147}, Nevanlinna analytical continuation (NAC)~\cite{PhysRevLett.126.056402} and its extension for bosonic systems~\cite{BNAC} and matrix-valued Green's functions (Carath\'{e}odory formalism~\cite{PhysRevB.104.165111}), etc. These methods can provide analytic forms of the Green's function and resolve complicated spectral functions over a wide range of frequencies with unprecedented accuracy. But they are not numerically stable, especially in the presence of noise. This deficiency largely restricts the applications of the interpolation-based methods. In addition to the above methods, there are still alternative routes, such as the non-negative least-squares method (NNLS)~\cite{Lawson1995}, non-negative Tikhonov method (NNT)~\cite{Tik1995,PhysRevB.107.085129}, sparse modeling (SpM)~\cite{PhysRevE.95.061302,PhysRevB.105.035139}, machine learning-aided methods~\cite{PhysRevLett.124.056401,PhysRevB.98.245101,PhysRevB.105.075112,PhysRevResearch.4.043082,Yao_2022,Arsenault_2017,PhysRevE.106.025312}, to name a few. However, to our knowledge, these methods have not yet been widely used to handle the realistic quantum Monte Carlo data.

Quite recently, two novel analytic continuation methods have been published. The first one is termed the Projection-Estimation-Semidefinite relaxation (PES) approach~\cite{PhysRevB.107.075151}. Just as its name suggests, the PES approach encompasses three important steps: (1) \emph{Causal projection}. The initial step involves projecting noisy Matsubara data onto a causal space, which is instrumental in mitigating the impact of unphysical noise and ensuring that the data adheres to the physical constraints (sum-rules) of the system. (2) \emph{Pole estimation}. Subsequent to the causal projection, the method employs the adaptive Antoulas-Anderson (AAA) algorithm~\cite{AAA,AAA_Lawson} to make a rough estimation about the locations of poles of the Matsubara Green's functions. The purpose of this step is to obtain a reasonable guess for the poles, which are important for the successive semidefinite relaxation optimization. (3) \emph{Semidefinite relaxation}. The final step involves a bi-level optimization algorithm to approximate Matsubara Green's functions using semidefinite relaxation. This algorithm effectively relaxes the rank-1 constraint on the semidefinite matrices, allowing for a more flexible and accurate fitting of the Matsubara data while enforcing the causality of the Green's functions. The PES approach is applicable to both fermionic and bosonic systems. It demonstrates improved accuracy and reliability in retrieving spectral features, especially in the presence of significant noise levels~\cite{PhysRevB.107.075151}. Furthermore, it does not require extended precision arithmetics, distinguishing it favorably from existing interpolation-based methods in the field~\cite{PhysRevB.87.245135,PhysRevB.93.075104,Vidberg1977,PhysRevD.96.036002,PhysRevB.61.5147,PhysRevLett.126.056402,BNAC,PhysRevB.104.165111}.

Another new analytic continuation approach is based on the minimal pole representation and the Prony's approximation (MPR)~\cite{PhysRevB.110.035154,PhysRevB.110.235131}. The MPR method also involves four essential steps. Initially, Matsubara Green's function on a finite interval of the imaginary axis is approximated by using the Prony's approximation~\cite{BEYLKIN200517,BEYLKIN2010131}, which approximates the Matsubara data as a sum of exponentials. Subsequently, this interval is mapped onto the unit circle via a holomorphic mapping. Then, the moments of the approximated function are numerically evaluated. The Prony's approximation is employed again to obtain a compact pole representation for the Matsubara Green's function. Finally, these poles are mapped back to the original domain, and the spectral function is evaluated. The MPR method offers a systematic and controlled approach to approximate the Matsubara Green's function within a predefined precision in terms of a minimal pole representation. It is generally applicable to the diagonal and off-diagonal Green's functions~\cite{PhysRevB.110.235131}.

Inspired by the PES and MPR methods~\cite{PhysRevB.110.035154,PhysRevB.107.075151,PhysRevB.110.235131}, we would like to introduce a new analytic continuation method in this paper. This method applies the barycentric rational functions to directly interpolate the original Matsubara Green's functions. The nodes and weights of the barycentric rational functions are determined by the AAA algorithm~\cite{AAA,AAA_Lawson}, providing a compact pole representation as well. If the Matsubara data is polluted with stochastic noise, the Prony's approximation is introduced to suppress the noise and guarantee numerical stability~\cite{BEYLKIN200517,BEYLKIN2010131}. This new method is called BarRat. Extensive tests on the BarRat method suggest that it works very well in most cases. It exhibits extremely high accuracy and efficiency, and is not very sensitive to data noise. These advantages make it a promising competitor to the popular MaxEnt method.

The rest of this paper is organized as follows. In Section~\ref{sec:method}, we introduce the basic ideas of the BarRat method, including the barycentric rational function, the AAA algorithm, and the Prony's approximation. They are the pivotal ingredients of the BarRat method. The implemented details are also elaborated in this section. Sections~\ref{sec:toys} and \ref{sec:examples} are devoted to the benchmarks of the BarRat method. Several toy models, including the diagonal and off-diagonal fermionic and bosonic Green's functions of continuous and discrete systems, are considered in Section~\ref{sec:toys}. Two concrete examples, namely Nambu Green's functions and self-energy functions, are handled in Section~\ref{sec:examples}. We discuss several important issues about the BarRat method, including its robustness with respect to noisy Matsubara data, data denoising by the Prony's approximation, size of input data, computational efficiency, and relations with the other analytic continuation methods in Section~\ref{sec:disc}. Finally, Section~\ref{sec:con} serves as a short conclusion. We look forward to further applications of the BarRat method in other research fields.

\section{Formalisms\label{sec:method}}

\subsection{Spectral representation}

In essence, the BarRat method should be classified as the interpolation-based analytic continuation method. Thus, it is not surprising that this method suits Matsubara Green's functions only. In this section, we would like to retrospect some basic knowledge about the spectral representation of Matsubara Green's functions at first.

Just as mentioned as before, in the context of quantum many-body systems, the Matsubara Green's function $G(i\omega_n)$ is bound to the spectral function $A(\omega)$ via the Laplace transformation [see Eq.~(\ref{eq:fredholm})]. For fermionic systems, the spectral function is positive definite, i.e., $A(\omega) > 0$. We have:
\begin{equation}
\label{eq:KA_f}
G(i\omega_n) = \int_{-\infty}^{\infty} d\omega~K(\omega_n, \omega) A(\omega),
\end{equation}
where $\omega_n = (2n + 1) \pi / \beta$, $n$ is a non-negative integer, and $\beta$ is the inverse temperature of the system ($\equiv 1/T$). The kernel function $K(\omega_n, \omega)$ is given by:
\begin{equation}
\label{eq:kernel_f}
K(\omega_n, \omega) = \frac{1}{i\omega_n - \omega}.
\end{equation}
For bosonic systems, the spectral function obeys the following restriction:
\begin{equation}
\text{sign}(\omega) A(\omega) \ge 0.
\end{equation}
Thus, it is more convenient to introduce a regulated spectral function $\tilde{A}(\omega) \equiv A(\omega)/\omega$. Clearly, $\tilde{A}(\omega)$ is positive definite. Now we have:
\begin{equation}
\label{eq:KA_b}
G(i\omega_n) = \int_{-\infty}^{\infty} d\omega~K(\omega_n, \omega) \tilde{A}(\omega),
\end{equation}
where $\omega_n = 2n\pi/\beta$. The kernel function $K(\omega_n, \omega)$ reads:
\begin{equation}
\label{eq:kernel_b}
K(\omega_n, \omega) = \frac{\omega}{i\omega_n - \omega}.
\end{equation}
Especially, $K(0,0) = -1$. Now let us consider a special case of bosonic systems. If the bosonic operators are Hermitian, $\tilde{A}(\omega)$ is an even function, and the limit of integral in Eq.~(\ref{eq:KA_b}) is reduced from $(-\infty,\infty)$ to $(0,\infty)$. So, Eq.~(\ref{eq:KA_b}) can be transformed into:
\begin{equation}
\label{eq:KA_h}
G(i\omega_n) = \int_{0}^{\infty} d\omega~K(\omega_n, \omega) \tilde{A}(\omega).
\end{equation}
The kernel function $K(\omega_n, \omega)$ becomes:
\begin{equation}
\label{eq:kernel_h}
K(\omega_n, \omega) = \frac{-2\omega^2}{\omega^2_n + \omega^2}.
\end{equation}
Especially, $K(0,0) = -2$.

Actually, the BarRat method doesn't rely on the spectral representation of Matsubara Green's function. In other words, it won't solve Eqs.~(\ref{eq:KA_f}), (\ref{eq:KA_b}), and (\ref{eq:KA_h}) directly. These equations can be used to reproduce $G(i\omega_n)$ once $A(\omega)$ or $\tilde{A}(\omega)$ is determined. In this work, we just employed them to synthesize trial Matsubara data. Please see Section~\ref{subsec:setup} for more details.

\subsection{Barycentric rational function approximation}

The rational function $r(z)$ is often used to establish an approximation for a function $f(z)$ on a real or complex domain:
\begin{equation}
\label{eq:rf}
r(z) = \frac{p(z)}{q(z)} = \frac{p_0 + p_1 z^1 + \cdots + p_n z^n}{q_0 + q_1 z^1 + \cdots + q_m z^m},
\end{equation}
where $z \in \mathbb{R}$ or $\mathbb{C}$, $p(z)$ and $q(z)$ are polynomials, and the degree of $r(z)$ is $N = n + m$. We note that the famous Pad\'{e} approximation belongs to the rational function approximation~\cite{PhysRevB.87.245135,PhysRevB.93.075104,Vidberg1977,PhysRevD.96.036002,PhysRevB.61.5147}. Actually, there are no limits on the forms of $p(z)$ and $q(z)$. Our objective is to find the best rational function approximation to $G(z)$. In the present work, we just adopt the barycentric quotient, instead of Eq.~(\ref{eq:rf}) to interpolate $G(z)$.

The barycentric formula takes the form of a quotient of two partial fractions~\cite{S0036144502417715,17M1132409},
\begin{equation}
\label{eq:rfa}
b(z) = \frac{n(z)}{d(z)}
     = \sum^m_{j=1} \frac{w_j f_j}{z - z_j}
     {\Huge/} \sum^m_{j=1} \frac{w_j}{z - z_j},
\end{equation}
where $m \ge 1$ is an integer, $z_1, \cdots, z_m$ are a set of complex distinct support points (i.e., ``nodes''), $f_1, \cdots, f_m$ are a set of complex data values, and $w_1, \cdots, w_m$ are a set of complex weights. Here, in order to distinguish from $p(z)$ and $q(z)$, we just let $n(z)$ and $d(z)$ stand for the partial fractions in the numerator and the denominator, respectively. Now let us introduce the node polynomial $l(z)$:
\begin{equation}
l(z) = \prod^{m}_{j=1} (z - z_j).
\end{equation}
It is a monic polynomial of degree $m$ with the set $z_1, \cdots, z_m$ as roots. If we define:
\begin{equation}
p(z) = l(z) n(z),
\end{equation}
and
\begin{equation}
q(z) = l(z) d(z),
\end{equation}
then both $p(z)$ and $q(z)$ are polynomials of degree at most $m - 1$. Thus, the barycentric quotient becomes~\cite{AAA}:
\begin{equation}
b(z) = \frac{n(z)}{d(z)} = \frac{l(z)n(z)}{l(z)d(z)} = \frac{p(z)}{q(z)}.
\end{equation}
This equation implies that $b(z)$ is a rational function as well.

A key aspect of barycentric quotient is its interpolatory property. According to Eq.~(\ref{eq:rfa}), at each point $z_j$ with $w_j \neq 0$, $b(z)$ becomes $\infty / \infty$. However, this singularity can be removed because $\lim_{z \to z_j} b(z) = f_j$. Thus, if the weights $w_1,~\cdots,~w_m$ are nonzero, Eq.~(\ref{eq:rfa}) provides a rational interpolant to the data $f_1,~\cdots,~f_m$ at $z_i,~\cdots,~z_m$. Now let us turn to the analytic continuation of Matsubara Green's function again. We just assume that the size of Matsubara data is exactly $m$, $f_j \equiv G(i\omega_j)$, $z_j \equiv i\omega_j$, and $j \in [1,m]$. Thus, once the weights $w_j$ are determined, we establish a barycentric quotient $b(z)$ to interpolate $G(z)$. Then, substituting $z$ with $\omega + i\eta$ in Eq.~(\ref{eq:rfa}), we can immediately get the retarded Green's function $G(\omega)$:
\begin{equation}
\label{eq:retarded}
G(\omega) = \lim_{\eta \to 0} b(\omega + i\eta),
\end{equation}
and the spectral function $A(\omega)$:
\begin{equation}
\label{eq:spectrum}
A(\omega) = -\frac{1}{\pi} \text{Im} G(\omega)
\end{equation}

\subsection{Adaptive Antoulas-Anderson algorithm}

So, the remaining problem is how to evaluate the weights $w_j$ in the barycentric rational function. In this work, we resort to the AAA algorithm, which is a fast and flexible method for near-best complex rational approximation~\cite{AAA,AAA_Lawson}. The AAA algorithm adaptively selects support points (i.e., $z_1,~\cdots,~z_m$) in a greedy manner, incrementally building the approximation degree one step at a time to avoid numerical instabilities. This algorithm ensures that the rational approximation is well-conditioned and can accurately capture the behavior of the target function, particularly in regions where it exhibits singularities or unbounded growth. Next, we will introduce the technical details about the AAA algorithm~\cite{AAA}.

\emph{Core AAA algorithm}. The AAA algorithm is actually an iterative approach. It begins with a finite sample set $Z \subseteq \mathbb{C}$ of $m \gg 1 $ points. The given function $f(z)$ is existing at least for all $z \in Z$. For each iteration $j = 1,~\cdots,~m$, the rational function approximation to $f(z)$ takes the barycentric form, i.e., $b_j(z)$. At step $j$, we first pick the next node $z_j$ by the greedy algorithm (see below). Then, we calculate the weights $w_1, w_2, \cdots, w_j$ by solving a linear least-squares problem (see below) over the remaining support points $Z^{(j)}$. Note that $Z^{(j)}$ forms a subset of support points,
\begin{equation}
\label{eq:Zj}
Z^{(j)} = Z \backslash \{z_1,~z_2,~\cdots,~z_j\}.
\end{equation}
Thus, at step $j$, we get the barycentric rational function $b_j(z)$, which generally interpolates $f_1,~f_2,~\cdots,~f_j$ at $z_1,~z_2,~\cdots,~z_j$. The AAA algorithm will terminate when the residual $||f(z) - b_j(z)||$ is sufficiently small. It is suggested that a default tolerance of $10^{-13}$ relative to the maximum of $|f(z)|$ is enough. Once the algorithm terminates, the barycentric rational approximation $b(z)$ to $f(z)$ is obtained, along with the poles, residues, and zeros of the interpolant (see below).

\emph{Greedy algorithm}. At step $j$ of the AAA algorithm, the next node $z_j$ must be chosen from $Z^{(j-1)}$ in a greedy manner. Specifically, we should go through every node in $Z^{(j-1)}$ and calculate the nonlinear residual $||f(z) - b_{j-1}(z)||$. At $z_j$, the nonlinear residual takes its maximum absolute value. In other words,
\begin{equation}
\label{eq:greedy}
z_j = \mathop{\arg \max}\limits_{z}\left|\left|f(z) - b_{j-1}(z)\right|\right|_{z \in Z^{(j-1)}}.
\end{equation}

\emph{Linear least-squares problem}. Our aim is an approximation $f(z) \approx b(z) = n(z) / d(z)$. Its linearized form becomes
\begin{equation}
f(z) d(z) \approx n(z).
\end{equation}
The weights $w_1,~w_2,~\cdots,~w_j$ in $b_j(z)$ are determined by solving the following least-squares problem:
\begin{equation}
\label{eq:lsp}
\mathop{\arg \min}\limits_{w_1,~\cdots,~w_j}\left|\left|f(z) - b_{j}(z)\right|\right|_{z \in Z^{(j-1)}},
\end{equation}
under the constraint $||w||_j = 1$. Here, $||\cdot||_{Z^{(j)}}$ is the discrete 2-norm over $Z^{(j)}$, $||\cdot||_j$ is the discrete 2-norm on $j-$vectors. Eq.~(\ref{eq:lsp}) is simplified to:
\begin{equation}
\label{eq:lsp_s}
\mathop{\arg \min}\limits_{w_1,~\cdots,~w_j}\left|\left| A^{(j)} w \right|\right|_{m - j},
\end{equation}
where $A^{(j)}$ is the $(m - j) \times j$ Loewner matrix~\cite{doi:10.1137/1.9780898718713}:
\begin{equation}
\label{eq:Aj}
A^{(j)} =
\left[\begin{array}{ccc}
\frac{F^{(j)}_1 - f_1}{Z^{(j)}_1 - z_1} & \cdots & \frac{F^{(j)}_1 - f_j}{Z^{(j)}_1 - z_j} \\
\vdots & \ddots & \vdots \\
\frac{F^{(j)}_{m-j} - f_1}{Z^{(j)}_{m-j} - z_1} & \cdots & \frac{F^{(j)}_{m-j} - f_j}{Z^{(j)}_{m-j} - z_j} \\
\end{array}\right],
\end{equation}
with $F^{(j)} \equiv f(Z^{(j)})$. Eq.~(\ref{eq:lsp_s}) is easily solved using the singular value decomposition of $A^{(j)}$. Actually, $w$ is the final right singular vector of $A^{(j)}$.

\emph{Poles and zeros of barycentric quotient}. In principle, the zeros of $d(z)$ are exactly the poles of $b(z)$. They can be computed by solving the following generalized eigenvalue problem~\cite{GK}:
\begin{equation}
\label{eq:eigen}
\left(\begin{array}{ccccc}
0 & w_1 & w_2 & \cdots & w_j \\
1 & z_1 &     &        & \\
1 &     & z_2 &        & \\
\vdots &&     & \ddots & \\
1 &     &     &        & z_j
\end{array}\right)
= \lambda
\left(\begin{array}{ccccc}
0&&&& \\
&1&&& \\
&&1&& \\
&&&1& \\
&&&&1 \\
\end{array}\right).
\end{equation}
The zeros of $n(z)$ are the zeros of $b(z)$ as well. They can be evaluated in a similar way by replacing $w_j$ with $w_j f_i$ in the above equation.

The AAA algorithm is robust. It avoids the pitfalls of exponential instabilities that can plague other rational approximation methods, particularly when approximating functions with poles or singularities. The AAA algorithm is flexible. It can be applied to a variety of domains, including intervals, disks, and more complex geometries, making it a competitive choice for rational approximation tasks~\cite{AAA}.

\subsection{Prony's approximation}

As mentioned before, realistic Matsubara Green's function data from quantum Monte Carlo simulations usually include non-trivial noise~\cite{gubernatis_kawashima_werner_2016,RevModPhys.83.349,RevModPhys.73.33}. Although the BarRat method (actually the AAA algorithm)~\cite{AAA,AAA_Lawson} exhibits excellent robustness with respect to noisy data, it may not respect analytic properties (Nevanlinna or Carath\'{e}odory structure) of the Matsubara Green's functions~\cite{PhysRevLett.126.056402,PhysRevB.104.165111}. Thus, if the noise level is remarkable, the BarRat method may yield some unphysical features in the spectral functions. In order to mitigate this problem, the Prony's approximation is adopted to suppress the possible noise and fluctuation in input data.

We assume the input Matsubara data consists of an odd number $m$ of Matsubara points $f_j$. They are uniformly spaced. The Prony's approximation just approximates the Matsubara data $f_j$ as a sum of exponentials for all $1 \le j \le m$~\cite{BEYLKIN200517,BEYLKIN2010131}:
\begin{equation}
\label{eq:prony}
\left|f_j - \sum^{K}_{i=1} p_i \gamma^j_i\right| \le \epsilon,
\end{equation}
where $\epsilon$ is a predefined tolerance ($\epsilon > 0$), $p_i$ denotes complex weights, and $\gamma_i$ means corresponding nodes. For Matsubara Green's functions, $K \propto \log{(1/\epsilon)}$, and only $K$ nodes in the Prony's approximation have weights $|p_i| > \epsilon$. These significant nodes $\gamma_i$ can be predetermined, such that the algorithm for finding weights $p_i$ is stable. Finally, an approximation of input Matsubara data is obtained. It requires a minimum number of nodes.

We would like to emphasize that the Prony's approximation is beneficial. Whereas it is neither compulsory nor the only choice. It is possible to be replaced with other denoising algorithms, such as the causal projections~\cite{PhysRevB.107.075151} and the matrix pencil method~\cite{56027}. Improved estimators for the quantum Monte Carlo algorithms~\cite{RevModPhys.83.349} and advanced representations for Matsubara Green's functions~\cite{PhysRevB.84.075145,PhysRevB.96.035147} might be helpful as well.

\begin{figure*}[ht]
\centering
\includegraphics[width=\textwidth]{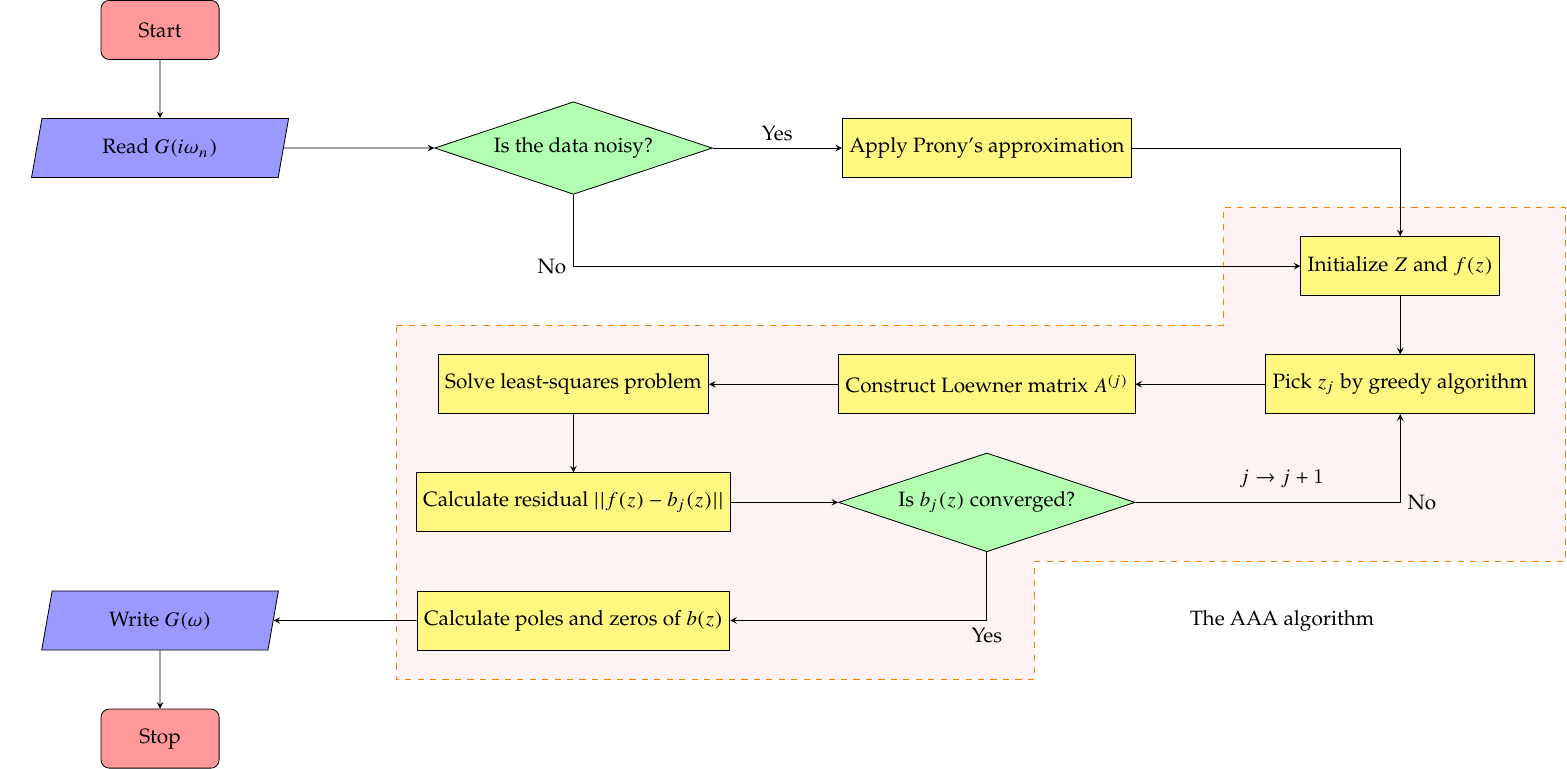}
\caption{Schematic workflow of the BarRat method as implemented in the \texttt{ACFlow} toolkit~\cite{Huang:2022,github}. The flowchart of the AAA algorithm is enclosed by a dashed line. \label{fig:flow}}
\end{figure*}

\subsection{Reference implementation}

A pedagogical implementation of the BarRat method is given in the \texttt{ACFlow} package, which is a full-fledged open-source analytic continuation toolkit~\cite{Huang:2022,github}. In addition to the BarRat method, the \texttt{ACFlow} toolkit also supports several other popular analytic continuation methods, such as the MaxEnt~\cite{PhysRevB.41.2380,JARRELL1996133,Bryan1990,PhysRevB.44.6011,PhysRevB.81.155107,PhysRevE.94.023303,PhysRevB.98.205102,PhysRevB.96.155128,PhysRevB.95.121104,PhysRevB.92.060509}, SAC~\cite{beach2004,PhysRevB.57.10287,PhysRevE.94.063308,PhysRevB.76.035115,PhysRevX.7.041072,SHAO20231}, SOM~\cite{PhysRevB.62.6317,PhysRevB.95.014102,KRIVENKO2019166,KRIVENKO2022108491}, SPX~\cite{PhysRevB.108.235143,PhysRevD.109.054508}, and NAC~\cite{PhysRevLett.126.056402} methods, etc. Therefore, the \texttt{ACFlow} toolkit provides a versatile platform to benchmark the performance of various analytic continuation methods.

The workflow of the BarRat method is illustrated in Fig.~\ref{fig:flow}. Next, we would like to explain some important steps. (1)~\emph{Apply Prony's approximation}. This step is optional. By solving Eq.~(\ref{eq:prony}), the weights $p_i$ and nodes $\gamma_i$ are determined. Then they are used to approximate the Matsubara data. (2)~\emph{Initialize $Z$ and $f(z)$}. This is the initial step of the AAA algorithm. In this step, we should use the Matsubara data to initialize $Z$ and $f(z)$. (3)~\emph{Pick $z_j$ by greedy algorithm}. Here is the entrance of the $j$-th iteration of the AAA algorithm. First of all, $Z^{(j-1)}$, which contains the unused support points, is constructed by using Eq.~(\ref{eq:Zj}). Then the next node $z_j$ is determined by solving Eq.~(\ref{eq:greedy}). This step is quite time-consuming, because we have to calculate the residual $||f(z) - b_{j-1}(z)||$ for every element in $Z^{(j-1)}$. (4)~\emph{Calculate Loewner matrix $A^{(j)}$}. It is defined in Eq.~(\ref{eq:Aj}). Actually, the calculation is simplified by using the Cauchy matrix~\cite{AAA}. (5)~\emph{Solve least-squares problem}. Here, we have to solve Eq.~(\ref{eq:lsp}) to get the weights $w_1,~w_2,~\cdots,~w_j$ in $b_j(z)$. Virtually, we perform singular value decomposition (SVD) for $A^{(j)}$ [i.e., $A^{(j)} = U \Sigma V^{*}$]. The last vector of $V$ is indeed $w$. (6)~\emph{Calculate residual $||f(z) - b_j(z)||$}. At first, the weights $w$ are used to construct the barycentric quotient $b_j(z)$ [see Eq.~(\ref{eq:rfa})]. Then we calculate the residual. (7)~\emph{Is $b_j(z)$ converged?} If the residual is larger than the predefined tolerance, we should increase $j$ by 1, and go back to step (3). (8)~\emph{Calculate poles and zeros of $b(z)$}. Now we obtain an optimal rational function approximation $b(z)$ for $f(z)$. We should analyze the zeros of $d(z)$ and $n(z)$ to get the poles and zeros of $b(z)$. This step involves solving a generalized eigenvalue problem [see Eq.(\ref{eq:eigen})]. (9)~\emph{Write $G(\omega)$}. Finally, the retarded Green's function $G(\omega)$ is calculated via Eq.~(\ref{eq:retarded}). Then it is used to evaluate $A(\omega)$ via Eq.~(\ref{eq:spectrum}).

In the current implementation~\cite{github}, several numerical issues need to be emphasized.
\begin{itemize}

\item The BarRat method only supports analytic continuation for Matsubara data. A relatively large number of data points are needed (see Section~\ref{subsec:size_of_data}). Otherwise, the AAA algorithm may yield oscillating results.

\item The workflow as depicted in Fig.~\ref{fig:flow} works quite well for continuous spectra. However, if the spectral functions are discrete, the AAA algorithm may struggle to obtain the correct weights $w_1,~w_2,~\cdots,~w_j$. Therefore, we adopt a slightly different approach. At first, we try to calculate the poles of the barycentric quotient $b(z)$. Next, $b(z)$ is expressed as the pole representation~\cite{PhysRevB.108.235143,PhysRevB.110.035154}:
\begin{equation}
b(z) = \sum^{N_p}_{p = 1} \frac{A_p}{z - z_p},
\end{equation}
where $N_p$ means the number of poles, $z_p$ and $A_p$ are positions and amplitudes of the $p$-th pole, respectively. We only retain the poles that are in the vicinity of the real axis, leading to the following equation:
\begin{equation}
b(z) \approx \sum^{N'_p}_{p = 1} \frac{A'_p}{z - z'_p},
\end{equation}
where $N'_p$ is the number of retained poles ($N'_p \le N_p$), $|\text{Im} z'_p| < \epsilon$~\cite{PhysRevB.107.075151}. Next, the Broyden-Fletcher-Goldfarb-Shanno (BFGS) algorithm~\cite{optimization_book} is used to optimize the amplitudes $A'_p$. Finally, the pole representation of $b(z)$ is used to calculate the retarded Green's function:
\begin{equation}
G(\omega) = \lim_{\eta \to 0} b(\omega + i\eta)
\approx \lim_{\eta \to 0} \sum^{N'_p}_{p = 1} \frac{A'_p}{\omega + i\eta - z'_p}
\end{equation}

\item Several algorithms that aim to find the best estimates of $K$, $p_i$, and $\gamma_i$ parameters [see Eq.~(\ref{eq:prony})] have been proposed in the literature. These algorithms, such as the matrix pencil method~\cite{56027}, estimation of signal parameters via rotational invariance techniques~\cite{32276}, and Prony's approximation~\cite{PhysRevB.110.035154}, can be understood as variants of Prony's method. In this work, we just adopted the Prony's approximation, which was used in Ref.~[\onlinecite{PhysRevB.110.035154}] as well. In the future, it should be replaced with more robust and efficient methods.

\end{itemize}

\section{Applications: Toy models\label{sec:toys}}

\begin{table*}[ht]
\centering
\begin{tabular}{lllll}
\hline
\hline
System & Test & Model & Feature & Section \\
\hline
~ & T$_{1}$ $\sim$ T$_{3}$ & Lorentzian model & Multiple broad peaks & \ref{sec:lorentz_f} \\
Fermionic & T$_{4}$ $\sim$ T$_{6}$ & Pole model & Multiple off-centered $\delta$ peaks & \ref{sec:pole_f} \\
Green's functions & T$_{7}$ & Gaussian model & Multiple broad peaks + big gap & \ref{sec:gap_f} \\
~ & T$_{8}$ & Resonance model & Sharp band edges + big gap + wide platform & \ref{sec:gap_f} \\
\hline
Matrix-valued & T$_{9}$ &  Gaussian model & Multiple broad peaks & \ref{sec:matrix_f} \\
Green's functions & T$_{10}$ & Pole model & Multiple off-centered $\delta$ peaks & \ref{sec:matrix_f} \\
\hline
Bosonic & T$_{11}$ & Optical conductivity  & Narrow Drude peak + broad interband transition peak & \ref{sec:current} \\
Green's functions & T$_{12}$ & Optical conductivity & Sharp mid-infrared peak + broad interband transition peak & \ref{sec:current} \\
\hline
\hline
\end{tabular}
\caption{Overview of the 12 test cases. The matrix-valued Green's functions are fermionic. Notice that all cases are designed to represent typical scenarios one would encounter in practice.~\label{tab:examples}}
\end{table*}

\begin{figure*}[ht]
\centering
\includegraphics[width=\textwidth]{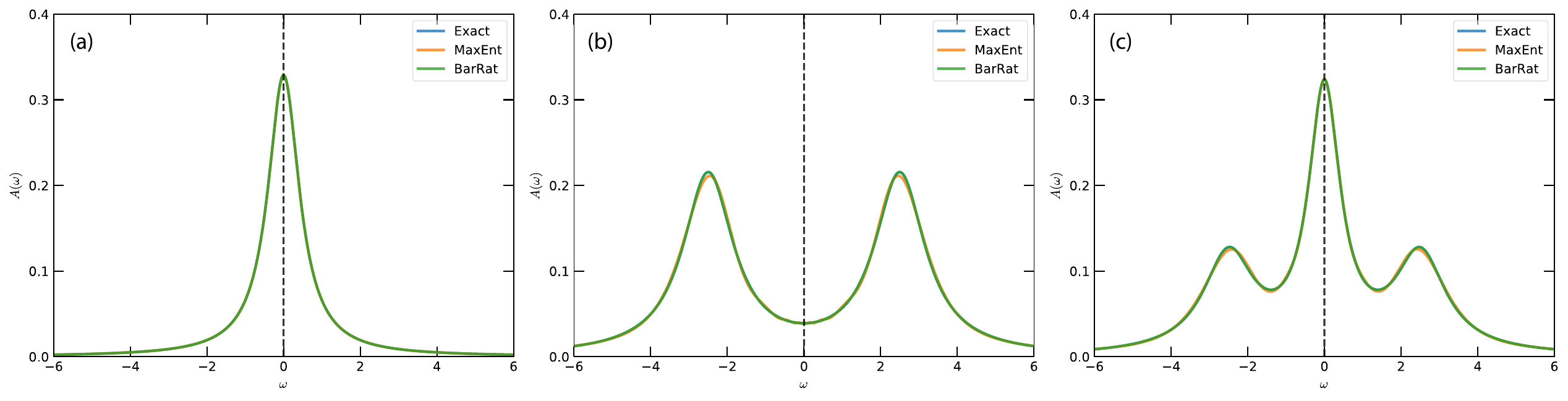}
\caption{Analytic continuations of fermionic Green's functions. (a) T$_{1}$: Single Lorentzian peak. (b) T$_{2}$: Two Lorentzian peaks. (c) T$_{3}$: Three Lorentzian peaks. The spectra shown in panel (a) are rescaled by a factor of 0.5 for a better view. The vertical lines denote the Fermi levels. See Section~\ref{sec:lorentz_f} for more technical details.~\label{fig:R01}}
\end{figure*}

\begin{figure*}[ht]
\centering
\includegraphics[width=\textwidth]{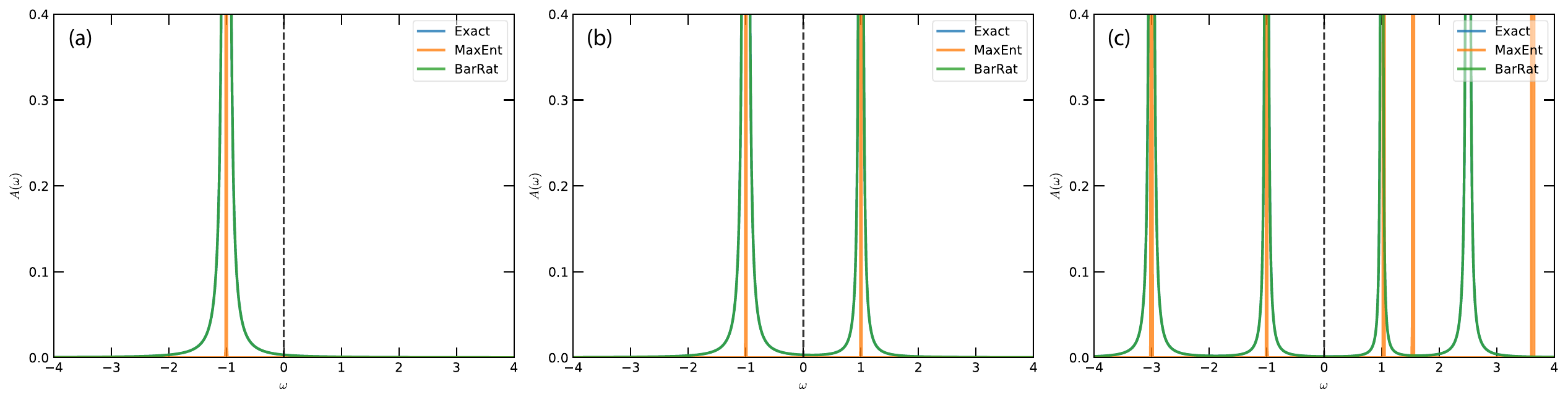}
\caption{Analytic continuations of fermionic Green's functions. (a) T$_{4}$: Single pole. (b) T$_{5}$: Two poles. (c) T$_{6}$: Four poles. The vertical lines denote the Fermi levels. See Section~\ref{sec:pole_f} for more technical details.~\label{fig:R02}}
\end{figure*}

\begin{figure}[ht]
\centering
\includegraphics[width=\columnwidth]{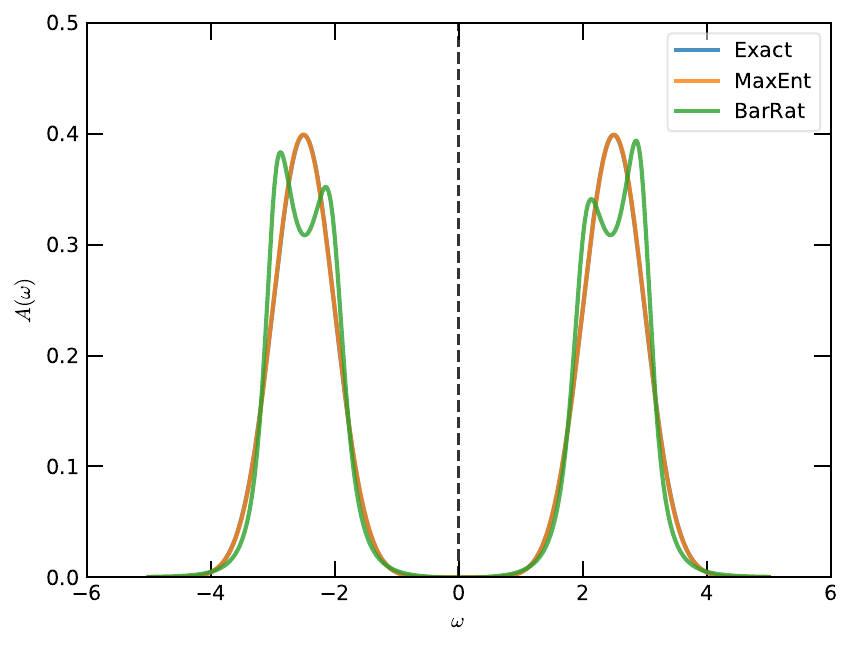}
\caption{Analytic continuation of fermionic Green's function (T$_{7}$: Two Gaussian peaks). The vertical line denotes the Fermi level. See Section~\ref{sec:gap_f} for more technical details.~\label{fig:X01}}
\end{figure}

\begin{figure}[ht]
\centering
\includegraphics[width=\columnwidth]{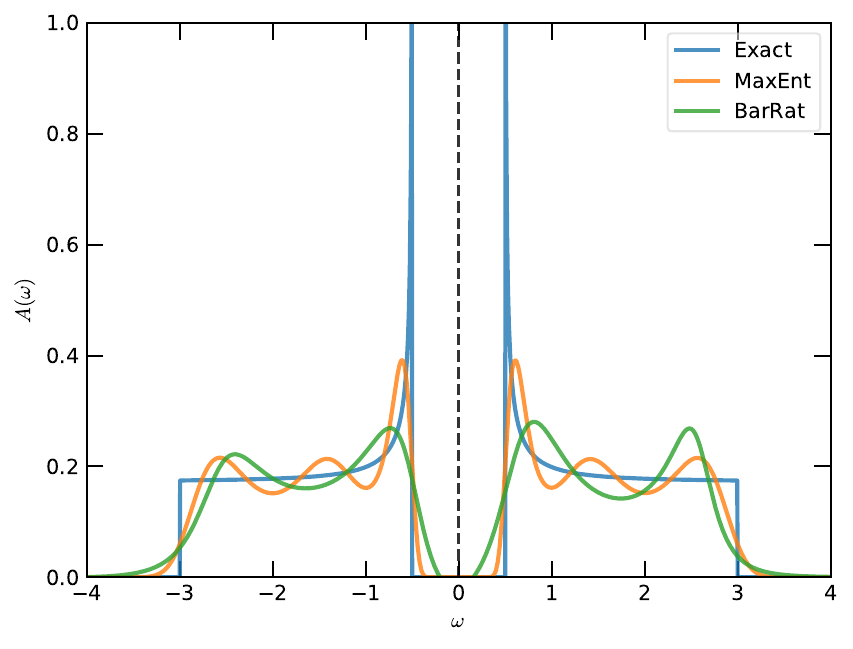}
\caption{Analytic continuation of fermionic Green's function (T$_{8}$: Resonance model). The vertical line denotes the Fermi level. See Section~\ref{sec:gap_f} for more technical details.~\label{fig:X04}}
\end{figure}

\begin{figure}[ht]
\centering
\includegraphics[width=\columnwidth]{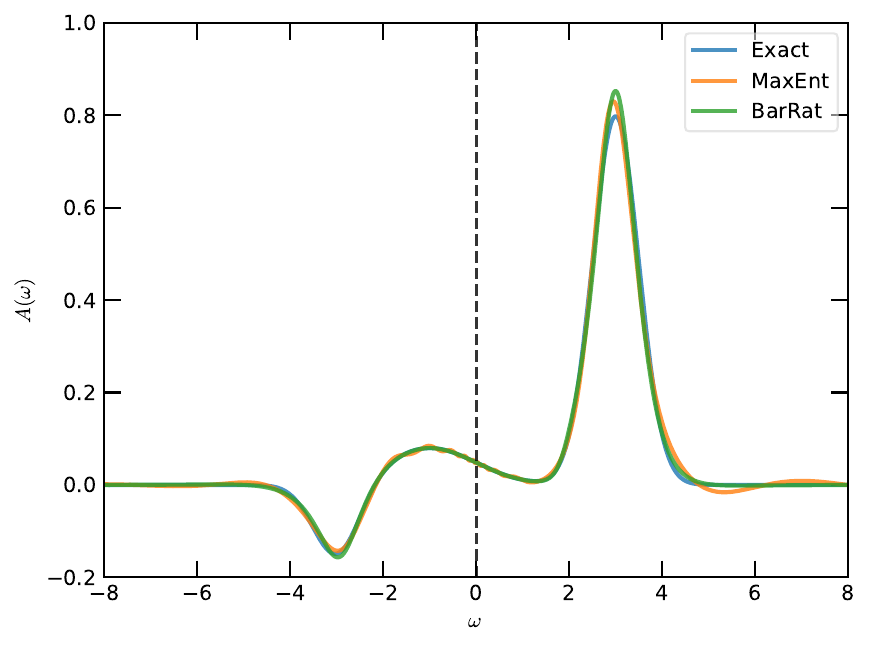}
\caption{Analytic continuation of off-diagonal Green's function (T$_{9}$: Three Gaussian peaks). The positive-negative entropy formalism is adopted in the MaxEnt method simulation. The vertical line denotes the Fermi level. See Section~\ref{sec:matrix_f} for more technical details.~\label{fig:R03}}
\end{figure}

\begin{figure}[ht]
\centering
\includegraphics[width=\columnwidth]{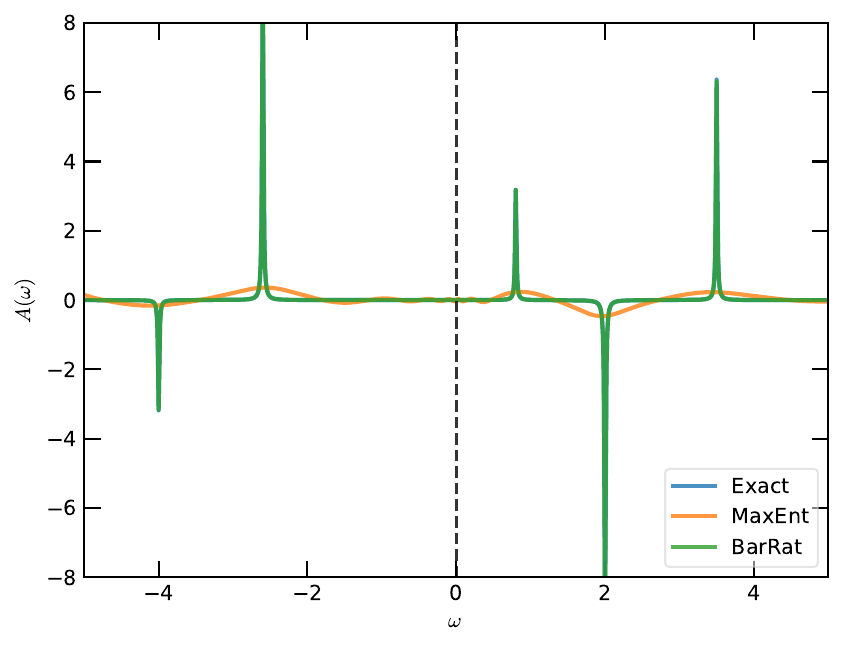}
\caption{Analytic continuation of off-diagonal Green's function (T$_{10}$: Five poles). The positive-negative entropy formalism is adopted in the MaxEnt method simulation. The vertical line denotes the Fermi level. See Section~\ref{sec:matrix_f} for more technical details.~\label{fig:R04}}
\end{figure}

\begin{figure}[ht]
\centering
\includegraphics[width=\columnwidth]{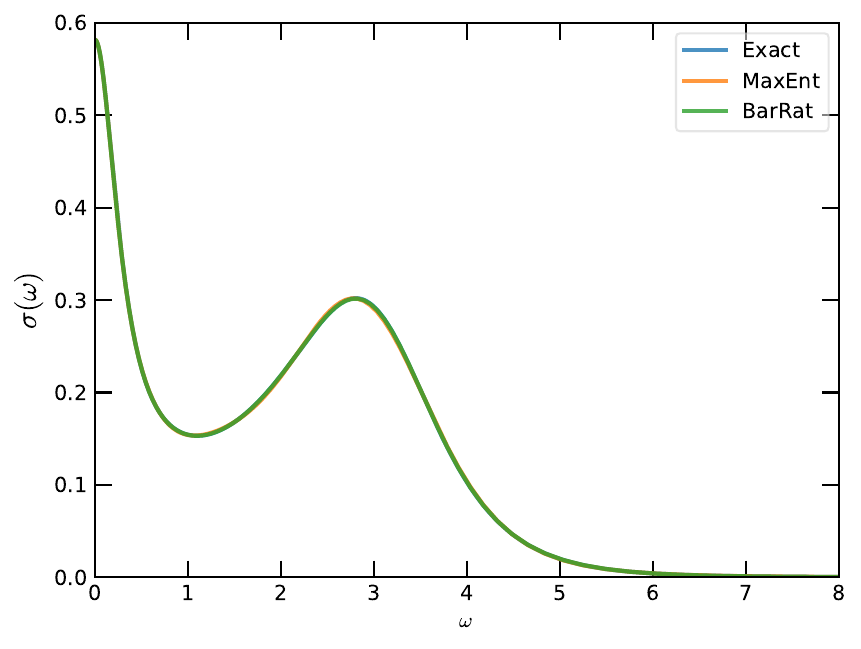}
\caption{Analytic continuation of current-current correlation function (Test T$_{11}$). See Section~\ref{sec:current} for more technical details.~\label{fig:R08}}
\end{figure}

\begin{figure}[ht]
\centering
\includegraphics[width=\columnwidth]{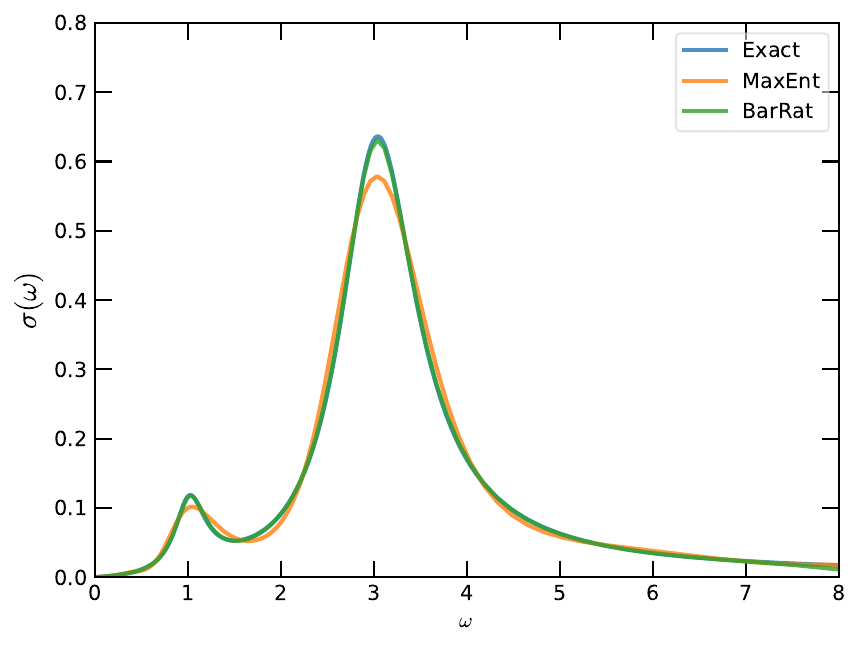}
\caption{Analytic continuation of current-current correlation function (Test T$_{12}$). See Section~\ref{sec:current} for more technical details.~\label{fig:R08N}}
\end{figure}

\subsection{Computational setup\label{subsec:setup}}

To assess the performance of the BarRat method, 12 test cases (designated as T$_{1}$ to T$_{12}$) have been constructed, covering fermionic correlators, bosonic correlators, and matrix-valued Green's functions. Their spectral functions are characteristic of those usually encountered in practical applications. A summary of these tests is provided in Table~\ref{tab:examples}.

To prepare the Matsubara data, we at first employ some analytic models to generate spectral functions $A_{\text{true}}(\omega)$. They are actually the exact solutions. These models include the parameterized Gaussian model, Lorentzian model, and pole model, etc. Next, the spectral functions $A_{\text{true}}(\omega)$ are used to calculate the Matsubara Green's functions $G_{\text{exact}}(i\omega_n)$ via Eqs.~(\ref{eq:KA_f})-(\ref{eq:kernel_h}). It is clear $G_{\text{exact}}(i\omega_n)$ is noiseless. Finally, additive Gaussian noise is introduced to the synthetic Green's function to mimic realistic conditions. The noise is incorporated using the following equation~\cite{PhysRevB.107.075151,PhysRevB.108.235143}:
\begin{equation}
\label{eq:noisy}
G_{\text{noisy}} = G_{\text{exact}} \left[1 + \delta \mathcal{N}_C(0, 1)\right],
\end{equation}
where $\mathcal{N}_C(0, 1)$ represents complex-valued Gaussian noise with zero mean and unit variance, and $\delta$ signifies the noise level of the data. Unless otherwise specified, $\delta$ is set to $10^{-4}$, the size of synthetic Matsubara data is $N_{\omega} = 100$, and the inverse temperature $\beta$ is set to 50.

The \texttt{ACFlow} toolkit was utilized for all analytic continuation calculations~\cite{Huang:2022}. In the present work, we employed two analytic continuation methods, namely the BarRat method and the MaxEnt method. The MaxEnt method is the most popular in this field~\cite{JARRELL1996133}. We would like to compare the spectra obtained by the two methods with the exact solutions. The BarRat method is almost parameter-free. By default, the Prony's approximation is disabled~\cite{PhysRevB.110.035154}. The MaxEnt method adopts the ``$\chi^2$kink'' algorithm~\cite{PhysRevE.94.023303} to identify the optimal regularization parameter $\alpha$, with its maximum value ranging from $10^9$ to $10^{15}$. The number of $\alpha$ parameters spans from 12 to 20. The default model is flat.

\subsection{Continuum spectra\label{sec:lorentz_f}}

In condensed matter physics, spectral functions are often continuous. We at first examine whether the BarRat method can resolve this type of spectral function. The exact spectral functions are constructed by a superposition of multiple Lorentzian functions (peaks). Its expression is as follows:
\begin{equation}
A_{\text{true}}(\omega) = \sum^{S}_{i = 1} \frac{1}{\pi} \frac{A_i \Gamma_i}{(\omega - \epsilon_i)^2+\Gamma^2_i}.
\end{equation}
Here, $S$ is the number of Lorentzian peaks, and $\epsilon_i$, $\Gamma_i$, and $A_i$ denote the center, broadening, and weight of the $i$-th Lorentzian peak, respectively. In this work, we consider three typical examples: (T$_{1}$) $S=1$, $\epsilon_1 = 0.0$, $\Gamma_1 = 0.5$, $A_1 = 0.5$. (T$_{2}$) $S=2$, $\epsilon_1 = -\epsilon_2 = 2.5$, $\Gamma_1 = \Gamma_2 = 0.8$, $A_1 = A_2 = 0.3$. (T$_{3}$) $S=3$, $\epsilon_1 = 0.0$, $\epsilon_2 = -\epsilon_3 = 2.5$, $\Gamma_1 = 0.5$, $\Gamma_2 = \Gamma_3 = 0.8$, $A_1 = 0.5$, $A_2 = A_3 = 0.3$. The computational results are displayed in Figure~\ref{fig:R01}. It can be observed that both the BarRat method and the MaxEnt method are capable of accurately reproducing the true spectral functions.

\subsection{Discrete spectra\label{sec:pole_f}}

In molecular systems (such as the Hubbard dimer), the spectral functions may consist of some discrete and sharp peaks~\cite{PhysRevB.107.075151,PhysRevB.104.165111}. Therefore, we would like to examine whether the BarRat method can resolve discrete spectra. The Matsubara Green's function is at first expressed in the form of pole expansion~\cite{PhysRevB.108.235143}:
\begin{equation}
\label{eq:matsubara_pole}
G(i\omega_n) = \sum^{S}_{i=1} \frac{A_i}{i\omega_n - \epsilon_i}.
\end{equation}
Here, $S$ represents the number of poles, and $A_i$ and $e_i$ represent the weight and position of the $i$-th pole, respectively. The retarded Green's function $G(\omega)$ can be easily calculated via Eq.~(\ref{eq:matsubara_pole}) through a variable substitution $i\omega_n \to \omega + i\eta$, where $\eta$ is a small parameter that measures the distance of the poles from the real axis. The spectral functions can be calculated via Eq.~(\ref{eq:spectrum}). In this work, we consider three typical cases ($\eta = 0.01$): (T$_{4}$) $S = 1$, $\epsilon_1 = -1.0$, $A_1 = 1.0$. (T$_{5}$) $S = 2$, $\epsilon_1 = -\epsilon_2 = 1.0$, $A_1 = 0.3$, $A_2 = 0.7$. (T$_{6}$) $S = 4$, $\epsilon_1 = -\epsilon_2 = 1.0$, $\epsilon_3 = 2.5$, $\epsilon_4 = -3.0$, $A_1 = 0.1$, $A_2 = 0.2$, $A_3 = 0.3$, $A_4 = 0.4$. The computational results are displayed in Figure~\ref{fig:R02}. We can see that the BarRat method can accurately resolve not only positions but also weights of the $\delta$-like peaks, even when they are far away from the Fermi level ($\omega = 0$). On the contrary, the MaxEnt method performs poorly. If the number of peaks is small and close to the Fermi level, the MaxEnt method can barely grasp positions of the peaks but neglects their broadening. If the number of peaks is large and the peaks are far from the Fermi level, the MaxEnt method usually fails. Similar results have been observed in previous publications~\cite{PhysRevLett.126.056402,PhysRevB.108.235143}. This is probably an inherent drawback of the MaxEnt method.

\subsection{Gapped systems\label{sec:gap_f}}

Next, let us turn to the gapped systems. We consider two spectral functions with large band gaps (Tests T$_7$ and T$_8$).

For Test T$_{7}$, the band edges are assumed to be smooth. Hence the spectral function is constructed using a superposition of two Gaussian peaks:
\begin{equation}
A_{\text{true}}(\omega) = \sum^{S}_{i = 1}
A_i
\exp{
\left[
-\frac{1}{2}
\left(\frac{\omega - \epsilon_i}{\Gamma_i}\right)^2
\right]
}.
\end{equation}
Here $S$ represents the number of Gaussian functions. $A_i$, $\epsilon_i$, and $\Gamma_i$ represent the weight, center, and broadening of the $i$-th Gaussian peak, respectively. The detailed parameters are $S = 2$, $A_1 = A_2 = 0.5$, $\epsilon_1 = -\epsilon_2 = 2.5$, $\Gamma_1 = \Gamma_2 = 0.5$. The computational results are illustrated in Figure~\ref{fig:X01}. In this test, the MaxEnt method can perfectly reproduce the true spectrum. The BarRat method performs slightly worse than the MaxEnt method. It can correctly resolve the band gap structure, but it yields additional peaks around $\omega = \pm 2.0$.

For Test T$_{8}$, we consider the scenario where the band edges are relatively sharp. This scenario is taken from Reference~[\onlinecite{beach2004}]. It concerns the analytic continuation of Matsubara Green's function of a BCS superconductor. The true spectral function reads:
\begin{equation}
\label{eq:bcs}
A_{\text{true}}(\omega) =
\left\{
    \begin{array}{lr}
    \frac{1}{W} \frac{|\omega|}{\sqrt{\omega^2 - \Delta^2}}, ~\quad & \text{if}~\Delta < |\omega| < W/2. \\
    0, & \text{otherwise}.
    \end{array}
\right.
\end{equation}
Here, $W$ denotes the total bandwidth, and $\Delta$ is used to control the gap's size (band gap = $2\Delta$). The detailed parameters are $W = 6.0$ and $\Delta = 0.5$. The spectrum is comprised of flat shoulders, steep peaks, and sharp gap edges. These distinctive features pose severe challenges to the existing analytic continuation methods~\cite{Huang:2022,PhysRevB.108.235143}. As is seen in Figure~\ref{fig:X04}, both the BarRat method and the MaxEnt method underestimate the energy gap and overestimate the bandwidth, introducing significantly unphysical oscillations in the plateau region and long tails in the high-frequency region. Based on our previous research results, perhaps only the SAC method and its variants (such as the SOM and SPX methods)~\cite{PhysRevB.108.235143,Huang:2022} combined with a constrained sampling algorithm can yield better results~\cite{SHAO20231,PhysRevE.94.063308}.

\subsection{Negative spectral weights\label{sec:matrix_f}}

In the preceding tests (T$_{1}$ $\sim$ T$_{8}$), all the spectral functions are positive definite. That is to say, $A(\omega) > 0$ and the sum-rule ($\int d\omega~A(\omega) = 1$) is fully satisfied. However, spectral functions are not necessarily positive definite. For instance, spectral functions of off-diagonal elements of matrix-valued Green's functions often do not fulfill positive definiteness~\cite{PhysRevB.98.205102,PhysRevB.96.155128}. Frequency-dependent transport coefficients, such as the Seebeck coefficient, Hall coefficient, Nernst coefficient, and so on, may also exhibit negative spectral weights~\cite{PhysRevB.95.121104,PhysRevB.92.060509}. The objective of this subsection is to examine whether the BarRat method can resolve non-positive definite spectral functions.

We consider two concrete tests (T$_{9}$ and T$_{10}$) again. For Test T$_{9}$, the spectral function is continuous, which is constructed by using the modified Gaussian functions:
\begin{equation}
A_{\text{true}}(\omega) = \sum^{S}_{i = 1}
\frac{A_i}{ \sqrt{2\pi} \Gamma_i}
\exp{
\left[-\frac{1}{2} \left(\frac{\omega - \epsilon_i}{\Gamma_i}\right)^2 \right]
}.
\end{equation}
The detailed parameters are $S = 3$, $A_1 = 0.5$, $-A_2 = A_3 = 0.1$, $\epsilon_1 = -\epsilon_2 = 3.0$, $\epsilon_3 = -1.0$, $\Gamma_1 = \Gamma_2 = 0.5$, and $\Gamma_3 = 1.0$. For Test T$_{10}$, the spectral function is discrete. The pole model is used again to build the Matsubara Green's function [see Eq.~(\ref{eq:matsubara_pole})]. The detailed parameters are $S = 5$, $\epsilon_1 = -4.0$, $\epsilon_2 = -0.26$, $\epsilon_3 =0.8$, $\epsilon_4 =2.0$, $\epsilon_5 =3.5$ and $A_1 = -0.1$, $A_2 = 0.3$, $A_3 = 0.1$, $A_4 = -0.3$, and $A_5 = 0.2$.

The BarRat method does not depend on the positive definiteness of the spectral function. Therefore, it can be used normally as long as the target spectral function is correctly set to be continuous or discrete. On the contrary, the MaxEnt method cannot directly resolve the non-positive definite spectral functions. There are some remedies, such as the auxiliary Green's function method~\cite{PhysRevB.90.041110,Tomczak_2007} and maximum quantum entropy method~\cite{PhysRevB.98.205102}. The simplest solution is possibly to extend the Shannon-Jaynes entropy to support the positive-negative entropy formalism~\cite{PhysRevE.94.023303}. In this work, we just adopted the positive-negative entropy approach. The analytic continuation results for Test T$_{9}$ and T$_{10}$ are illustrated in Figs.~\ref{fig:R03} and \ref{fig:R04}, respectively. We find the BarRat method is fully capable of dealing with non-positive definite spectral functions. It works quite well no matter whether the spectral function is continuous or discrete. The MaxEnt method supplemented by the positive-negative entropy formalism can effectively resolve continuous spectrum. However, for a discrete spectrum, this method can roughly identify the locations of the peaks but fails to reproduce their widths and weights. It tends to yield a smooth and continuous spectrum.

\subsection{Bosonic systems\label{sec:current}}

Next, we would like to concentrate on the analytic continuation of bosonic systems. In this subsection, we will show how to extract optical conductivity $\sigma(\omega)$ from the current-current correlation function $\Pi(i\omega_n)$ by using the BarRat method.

In the imaginary time axis, the current-current correlation function $\Pi(\tau)$ reads:
\begin{equation}
\Pi(\tau) = \frac{1}{3N} \langle \mathbf{j}(\tau) \cdot \mathbf{j}(0) \rangle,
\end{equation}
where $N$ is the number of sites, $\mathbf{j}$ is the current operator, and $\langle ... \rangle$ means the thermodynamic average~\cite{JARRELL1996133}. $\Pi(\tau)$ is a bosonic correlator. Since the BarRat method needs Matsubara data as input, we should convert $\Pi(\tau)$ to $\Pi(i\omega_n)$ via Fourier transformation in realistic simulations. The corresponding spectrum is the frequency-dependent optical conductivity $\sigma(\omega)$. In principle, $\sigma(\omega)$ is an even function, i.e., $\sigma(\omega) = \sigma(-\omega)$. The relation between $\Pi(i\omega_n)$ and $\sigma(\omega)$ reads:
\begin{equation}
\label{eq:current}
\Pi(i\omega_n) = \int^{+\infty}_{0} d\omega~
    K(\omega_n,\omega) \sigma(\omega).
\end{equation}
The kernel $K(\omega_n,\omega)$ is already defined in Eq.~(\ref{eq:kernel_h}). So, once the analytic expression of $\sigma(\omega)$ is known, then $\Pi(i\omega)$ can be generated by Eq.~(\ref{eq:current}) and Eq.~(\ref{eq:kernel_h}).

Here, we consider two individual models. The first model (T$_{11}$) is borrowed from Ref.~[\onlinecite{PhysRevB.82.165125}]. It reads:
\begin{equation}
\label{eq:optic}
\sigma(\omega) = \frac{T_1(\omega) + T_2(\omega) + T_3(\omega)}{1 + (\omega/\gamma_3)^6},
\end{equation}
and
\begin{eqnarray}
T_1(\omega) &=& \frac{\alpha_1}{1 + (\omega/\gamma_1)^2},\nonumber\\
T_2(\omega) &=& \frac{\alpha_2}{1 + [(\omega - \epsilon)/\gamma_2]^2},\nonumber\\
T_3(\omega) &=& \frac{\alpha_2}{1 + [(\omega + \epsilon)/\gamma_2]^2},
\end{eqnarray}
where $\epsilon$, $\alpha_i$, and $\gamma_i$ are adjustable parameters. Their values are $\alpha_1 = 0.3$, $\alpha_2 = 0.2$, $\gamma_1 = 0.3$, $\gamma_2 = 1.2$, $\gamma_3 = 4.0$, and $\epsilon = 3.0$. This spectrum manifests two peaks in the positive half-axis. The narrow one at $\omega = 0.0$ is called the Drude peak, which signals a metallic state. Another broad hump is at approximately $\omega = \epsilon$, which is usually from the contributions of interband transitions~\cite{RevModPhys.83.471}. The second model (Test T$_{12}$) reads:
\begin{equation}
\sigma(\omega) = \omega^{\alpha} \sum^{S}_{i=1}
\frac{1}{\pi} \frac{A_i \gamma_i}{(\omega - \epsilon_i)^2 + \gamma^2_i}.
\end{equation}
It is a variation of the Lorentzian model. The detailed parameters are $S = 2$, $\alpha = 0.5$, $A_1 = 0.1$, $A_2 = 0.5$, $\gamma_1 = 0.2$, $\gamma_2 = 0.5$, $\epsilon_1 = 1.0$, and $\epsilon_2 = 3.0$. In this model, the Drude peak at $\omega = 0.0$ disappears, the peak from interband transitions shifts to $\omega = \epsilon_2$, and a small satellite peak appears at $\omega = \epsilon_1$. This set of parameters corresponds to an insulating state.

The analytic continuation results of the two models are depicted in Figures~\ref{fig:R08} and~\ref{fig:R08N}. As can be seen from the figures, for the metallic state (Test T$_{11}$), both the BarRat method and the MaxEnt method can accurately reproduce the true $\sigma(\omega)$. For the insulating state (Test T$_{12}$), the BarRat method performs very well, once again perfectly reproducing the true optical conductivity. However, the MaxEnt method introduces visible deviations in the range of $1.0 < \omega < 3.0$.

\section{Applications: Realistic examples\label{sec:examples}}

\begin{figure}
\centering
\includegraphics[width=\columnwidth]{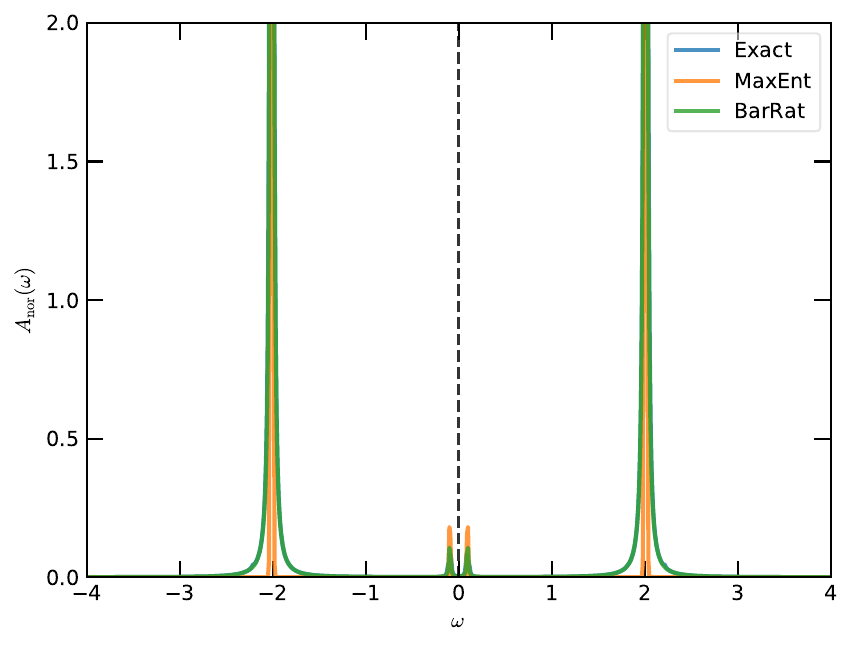}
\caption{Analytic continuations of Nambu Green's function (normal part) corresponding to the Anderson impurity model with a $s$-wave pairing bath [see Eq.~(\ref{eq:H_1band1bath})]. The parameters for the model Hamiltonian are $U=4.0$, $\beta=10.0$, $\Delta=0.1$, and $V=0.1$. \label{fig:Nambu_GW}}
\end{figure}

\begin{figure}
\centering
\includegraphics[width=\columnwidth]{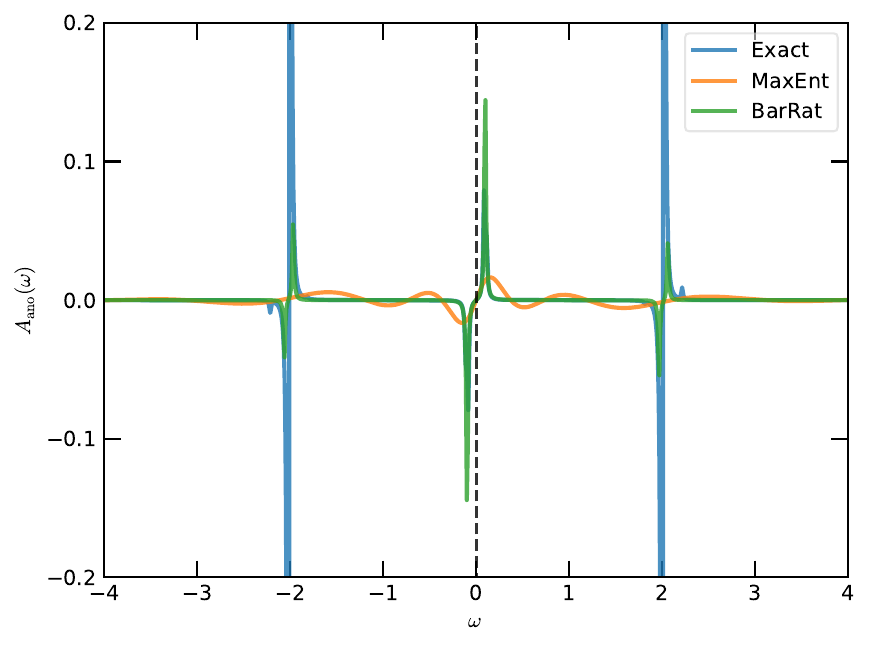}
\caption{Analytic continuations of Nambu Green's function (anomalous part) corresponding to the Anderson impurity model with a $s$-wave pairing bath [see Eq.~(\ref{eq:H_1band1bath})]. The parameters for the model Hamiltonian are $U=4.0$, $\beta=10.0$, $\Delta=0.1$, and $V=0.1$. \label{fig:Nambu_FW}}
\end{figure}

\begin{figure*}
\centering
\includegraphics[width=\textwidth]{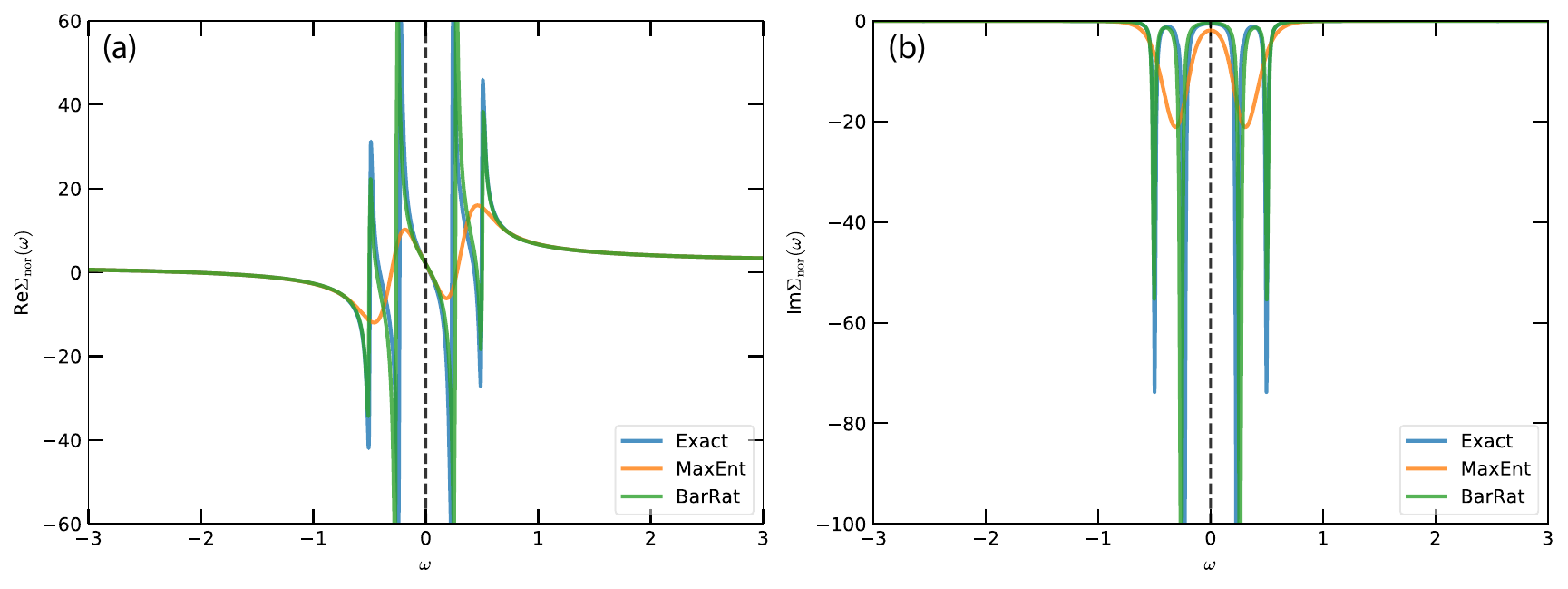}
\caption{Analytic continuations of Nambu self-energy function (normal part) corresponding to the Anderson impurity model with a $s$-wave pairing bath [see Eq.~(\ref{eq:H_1band1bath})]. The parameters for the model Hamiltonian are $U=4.0$, $\beta=10.0$, $\Delta=0.25$, and $V=0.2$. \label{fig:Nambu_SW1}}
\end{figure*}

\begin{figure*}
\centering
\includegraphics[width=\textwidth]{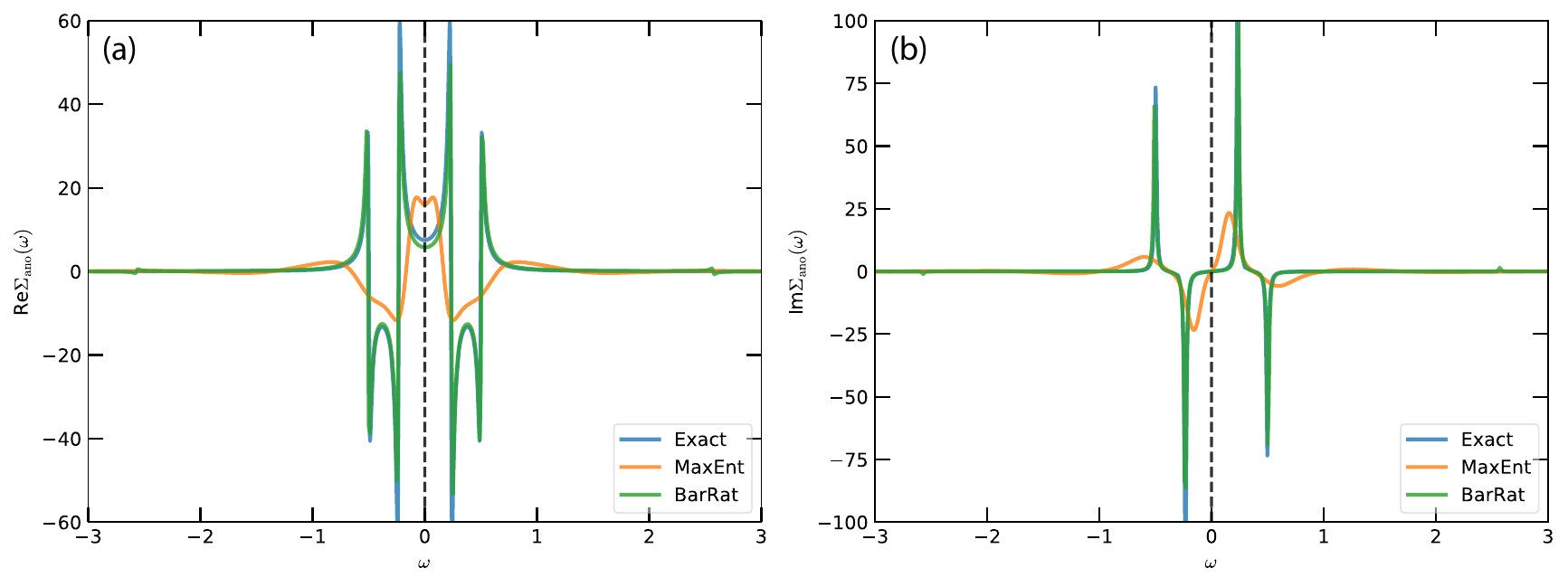}
\caption{Analytic continuations of Nambu self-energy function (anomalous part) corresponding to the Anderson impurity model with a $s$-wave pairing bath [see Eq.~(\ref{eq:H_1band1bath})]. The parameters for the model Hamiltonian are $U=4.0$, $\beta=10.0$, $\Delta=0.25$, and $V=0.2$. \label{fig:Nambu_SW2}}
\end{figure*}

In this section, we would like to apply the BarRat method to some realistic examples. Specifically, we will provide benchmarks to verify whether the BarRat method can handle the Nambu Green's and self-energy functions. They are representative matrix-valued correlation functions with nonzero off-diagonal elements.

\subsection{Model}

In the Gor'kov-Nambu formalism for the superconducting states, the self-energy function $\mathbf{\Sigma}$ is a matrix~\cite{PhysRev.117.648,Gorkov1958}. Taking a simple intra-orbital $s$-wave spin singlet pairing as an example, $\mathbf{\Sigma}$ is a $2 \times 2 $ matrix, with the normal part being the diagonal elements ($\Sigma_{\text{nor}} = \mathbf{\Sigma}_{11}$) and the anomalous part the off-diagonal elements ($\Sigma_{\text{ano}} = \mathbf{\Sigma}_{12}$). The spectral function of the anomalous self-energy function $\Sigma_{\text{ano}}(i\omega_n)$ is not positive definite, with multiple sign changes along the real axis. Quite recently, Yue \emph{et al.} proposed a novel method to perform analytic continuation for the anomalous self-energy function~\cite{PhysRevB.108.L220503}. They introduced an auxiliary self-energy function as a simple linear combination of normal and anomalous self-energy functions. This auxiliary function is proven to be positive definite. Then it is treated by the traditional MaxEnt method~\cite{JARRELL1996133}.

Here, we consider an Anderson impurity model with a single bath site with $s$-wave pairing \cite{PhysRevB.108.L220503}. The Hamiltonian reads:
\begin{align}
\label{eq:H_1band1bath}
H=&\, Un_{d\uparrow}n_{d\downarrow}-\mu(n_{d\uparrow}+n_{d\downarrow})+\epsilon_{b}c_{\uparrow}^{\dagger}c_{\uparrow}+\epsilon_{b}c_{\downarrow}^{\dagger}c_{\downarrow}\nonumber\\
&+\Delta(c_{\uparrow}c_{\downarrow}+h.c.)+(Vd_{\uparrow}^{\dagger}c_{\uparrow}+Vd_{\downarrow}^{\dagger}c_{\downarrow}+h.c.),
\end{align}
where $d$ and $d^{\dagger}$ are the impurity operators, $c$ and $c^{\dagger}$ are the bath operators, and $n_d$ and $n_c$ are the occupancy operators for impurity and bath, respectively. The other parameters are explained as follows: $U$ is the on-site interaction, $\mu$ the chemical potential, $V$ the hybridization parameter, $\epsilon_b$ the bath energy level, and $\Delta$ the pairing field. In this work, we choose the bath energy $\epsilon_b = 0$ and $\mu=0.5U$, which corresponds to a half-filling system. This Hamiltonian can be easily solved by the exact diagonalization (ED) method, as the total number of eigenstates is only 16. In the Nambu formalism, the Nambu spinor wave function $\psi$ is introduced:
\begin{equation}
\psi\equiv\left[\begin{array}{cc} c_{\uparrow}, & c_{\downarrow}^{\dagger}\end{array}\right]^{T}.
\end{equation}
The normal and anomalous Green's functions ($G_{\text{nor}} \equiv [\mathbf{G}]_{ii}$, $G_{\text{ano}} \equiv [\mathbf{G}]_{i \ne j}$) on the real or Matsubara frequency axis can be calculated by using the Lehmann representation~\cite{many_body_book,many_body_book_2016}:
\begin{equation}
[\mathbf{G}(z)]_{mn} =\frac{1}{Z} \sum_{i,j=1}^{16}
    \frac{e^{-\beta E_{\Gamma_{i}}}+e^{-\beta E_{\Gamma_{j}}}}{z-E_{\Gamma_{j}}+E_{\Gamma_{i}}}
    \langle\Gamma_{i}|\psi_{m}|\Gamma_{j}\rangle
    \langle\Gamma_{j}|\psi_{n}^{\dagger}|\Gamma_{i}\rangle,
\end{equation}
where $z = \omega + i\eta$ or $i\omega_n$, $Z$ is the partition function ($Z=\sum_{j=1}^{16}e^{-\beta E_{\Gamma_{j}}}$), $|\Gamma_i \rangle$ is the $i$-th eigenstate, and $E_{\Gamma_i}$ is the corresponding $i$-th eigenvalue. The Nambu self-energy function ${\bf \Sigma}$ is obtained by applying Dyson's equation in the Nambu formalism:
\begin{equation}
\mathbf{\Sigma}(z) \equiv
  \left[
    \begin{array}{cc}
        \Sigma_{\text{nor}}(z) & \Sigma_{\text{ano}}(z)  \\
        \Sigma_{\text{ano}}(z) & -\Sigma_{\text{nor}}(-z)
    \end{array}
  \right]
  = z\mathbf{I}-(\epsilon_{b}-\mu)\sigma_{3} - \mathbf{G}^{-1}(z),
\end{equation}
where $\mathbf{I}$ is the identity matrix and $\sigma_3$ is the Pauli matrix.

\subsection{Nambu Green's functions\label{sec:nambu_g}}

We at first consider the Nambu Green's functions. Note that we treat this matrix-valued function element by element, instead of as a whole. In other words, we perform analytic continuations for $G_{\text{nor}}(i\omega_n)$ and $G_{\text{ano}}(i\omega_n)$ separately. When the BarRat method is employed, we assume \emph{a priori} that the spectral function is discrete. If the MaxEnt method is used, then the positive-negative entropy approach~\cite{PhysRevE.94.023303} is utilized to handle the anomalous part. The calculated results are illustrated in Figures~\ref{fig:Nambu_GW} and~\ref{fig:Nambu_FW}. For analytic continuation of $G_{\text{nor}}(i\omega_n)$, the BarRat method perfectly reproduces the exact $A_{\text{nor}}(\omega)$. But the MaxEnt method, although capable of capturing the positions of all peaks, underestimates the width of the peaks at $\omega = \pm 2.0$. For analytic continuation of $G_{\text{ano}}(i\omega_n)$, the BarRat method still performs very well. Especially, it captures the antisymmetry of the spectral function $A_{\text{ano}}(\omega)$ near the Fermi level. It also accurately identifies the pole-like structures at $\omega = \pm 2.0$. As a comparison, the spectrum obtained by the MaxEnt method is generally an oscillatory curve, from which it is quite difficult to infer any valuable information.

\subsection{Nambu self-energy functions}

Next, we turn to the analytic continuation of $\Sigma_{\text{nor}}(i\omega_n)$ and $\Sigma_{\text{ano}}(i\omega_n)$. When $\omega_n$ goes to infinity, $\Sigma(i\omega_n)$ approaches $\Sigma_{\text{HF}}$ (a constant Hartree term), instead of $1/i\omega_n$. Thus, we should subtract the Hartree term  from $\Sigma(i\omega_n)$ in advance:
\begin{equation}
\tilde{\Sigma}(i\omega_n) = \Sigma(i\omega_n) - \Sigma_{\text{HF}}.
\end{equation}
Now the asymptotic behavior of $\tilde{\Sigma}(i\omega_n)$ is correct. Then, the analytic continuation of $\tilde{\Sigma}(i\omega_n)$ is performed to obtain $\tilde{\Sigma}(\omega)$, after which the Hartree term is added back:
\begin{equation}
\Sigma(\omega) = \tilde{\Sigma}(\omega) + \Sigma_{\text{HF}}.
\end{equation}
When applying the BarRat method, it is assumed that the spectral functions are discrete once again. As for the MaxEnt method, the positive-negative entropy approach~\cite{PhysRevE.94.023303} is used to handle the off-diagonal elements (i.e., the anomalous self-energy function). The analytic continuation results are shown in Figures~\ref{fig:Nambu_SW1} and~\ref{fig:Nambu_SW2}. It is remarkable that the BarRat method can fully reproduce all characteristics of the spectral functions, including the positions, intensities, and symmetries of the peaks, etc. The MaxEnt method can only provide a smooth envelope. Although it can roughly infer where the features are located, the details of the peaks are completely eliminated.

It is particularly worth emphasizing that we did not benchmark the auxiliary self-energy function method in the present work. The method proposed by Yue \emph{et al.} for analytic continuation of the Nambu self-energy function strictly ensures the positive definiteness of the spectral function of the auxiliary self-energy function~\cite{PhysRevB.108.L220503}. In a successive work, it would be interesting to test whether the BarRat method combined with the auxiliary self-energy function method is applicable to analytic continuations of the anomalous self-energy functions.

\section{Discussions\label{sec:disc}}

\begin{figure*}
\centering
\includegraphics[width=\textwidth]{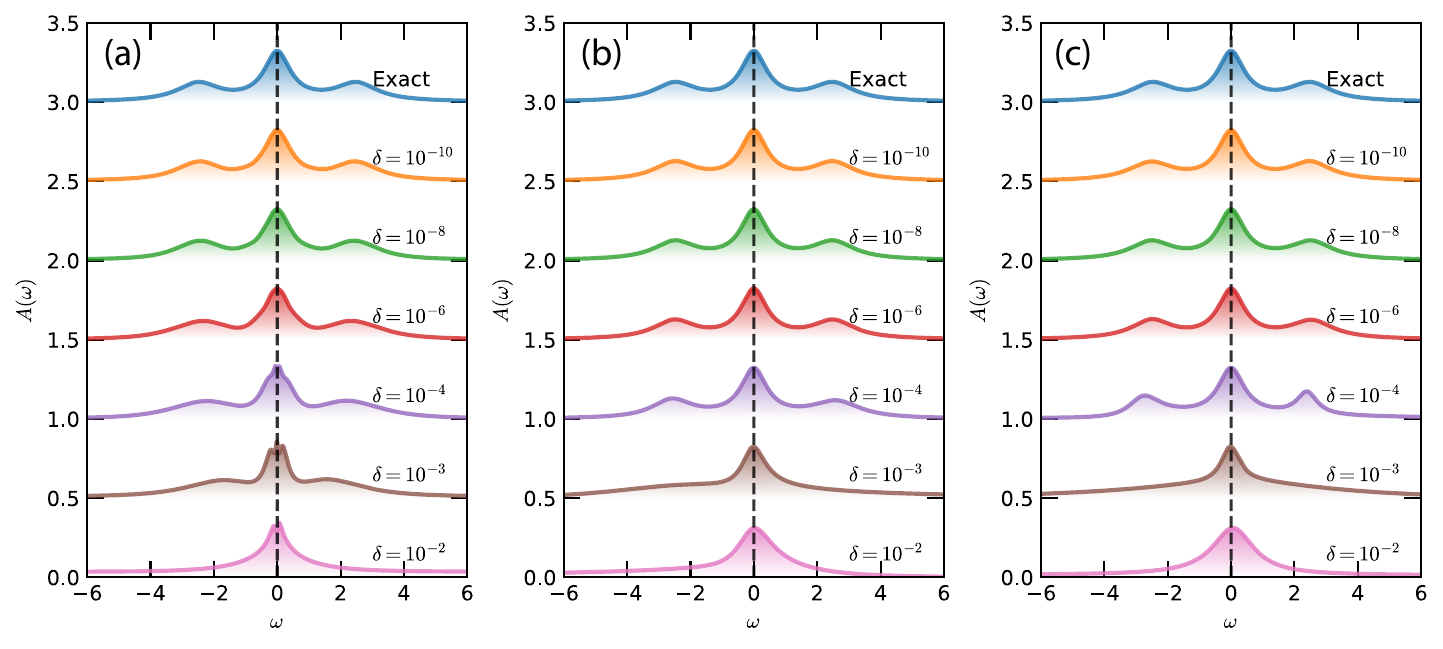}
\caption{Robustness of the BarRat method with respect to noisy Matsubara data for fermionic Green's functions. (a) Results by the MaxEnt method. (b) Results by the BarRat method. (c) Results by the BarRat method (the Prony's approximation is enabled for denoising). The vertical dashed lines denote the Fermi level. The noise level of input Matsubara data is controlled by the $\delta$ parameter [see Eq.~(\ref{eq:noisy})]. When the Prony's approximation is used, the denoising parameter $\epsilon$ is set to $\delta$ [see Eq.~(\ref{eq:prony})]. The benchmark data is taken from Test T$_{3}$. See Section~\ref{sec:lorentz_f} for more details about the test.~\label{fig:R01N}}
\end{figure*}

\begin{figure}
\centering
\includegraphics[width=\columnwidth]{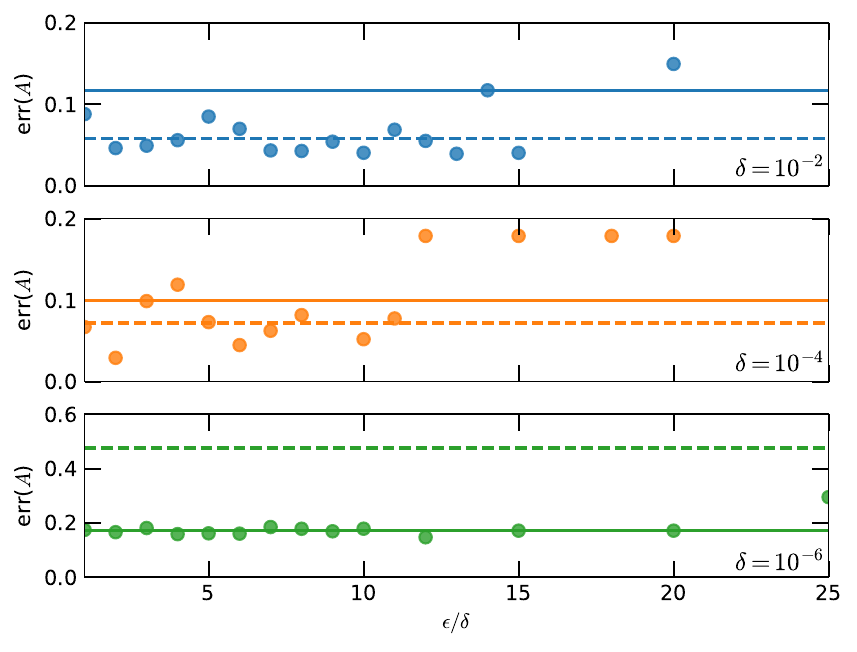}
\caption{Data denoising by the Prony's approximation. The input Matsubara data is taken from Test T$_{3}$. The noise level is controlled by the $\delta$ parameter [see Eq.~(\ref{eq:noisy})]. The solid and dashed lines denote the error values by the BarRat and MaxEnt methods, respectively. The filled circles are the error values by the BarRat method + Prony's approximation. They depend on the $\epsilon$ parameter that dominates the Prony's approximation [see Eq.~(\ref{eq:prony})]. \label{fig:denoise}}
\end{figure}

\begin{figure}
\centering
\includegraphics[width=\columnwidth]{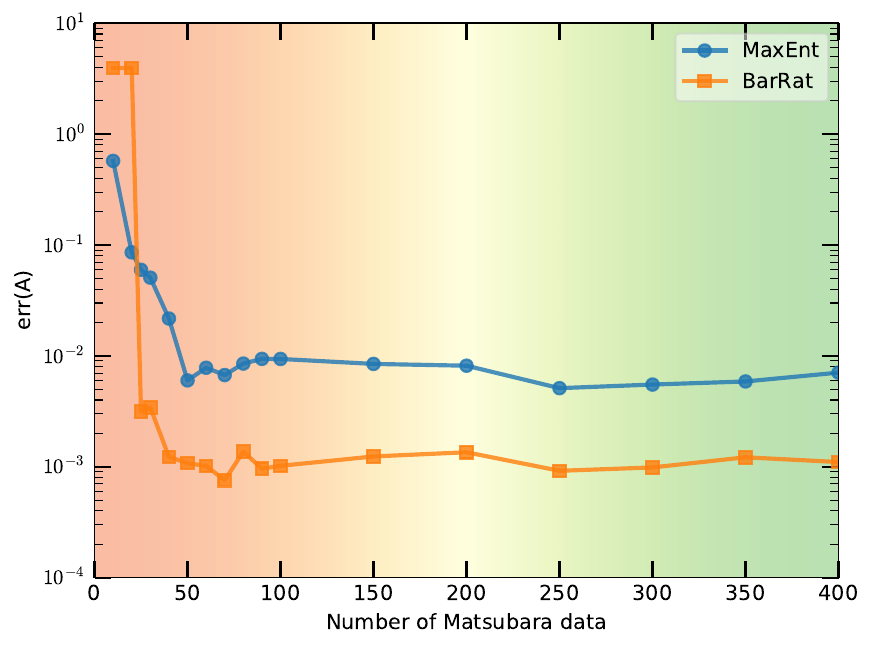}
\caption{Errors due to inadequate input data for the MaxEnt and BarRat methods. The input Matsubara data is taken from Test T$_{3}$. The noise level $\delta$ is fixed to $10^{-10}$. \label{fig:dataset}}
\end{figure}

In this section we will further discuss some essential issues about the BarRat method, including noise resistance, effect of denoising by the Prony's approximation, optimal size of the input dataset, efficiency, and the relationship with the other analytic continuation methods, etc.

\subsection{Noise tolerance}

We firstly benchmark the noise resistance of the BarRat method, which is a crucial evaluation metric for analytic continuation techniques. As many analytic continuation methods are highly efficient and accurate, their applications are limited due to their sensitivities to noise (such as the PA method~\cite{PhysRevB.87.245135,PhysRevB.93.075104,Vidberg1977,PhysRevD.96.036002,PhysRevB.61.5147} and the NAC method~\cite{PhysRevLett.126.056402,BNAC,PhysRevB.104.165111}). Although the MaxEnt method does not excel in computational accuracy, its robustness with respect to noisy data makes it one of the most popular analytic continuation methods~\cite{JARRELL1996133}.

Here we adopt Test T$_{3}$ to evaluate the noise resistance of the BarRat method. Its spectrum exhibits a three-peak structure, which is quite typical for correlated electron models (such as the single-band Hubbard model)~\cite{RevModPhys.68.13,RevModPhys.78.865}. In the preceding tests, the noise level $\delta$ for the synthetic Matsubara data is fixed to $10^{-4}$. Now $\delta$ varies from $10^{-10}$ to $10^{-2}$. We take two scenarios into consideration: (1) Only the BarRat method is used. (2) The BarRat method is augmented by the Prony's approximation. Figure~\ref{fig:R01N} depicts the simulated results. The results from the MaxEnt method are also shown in this figure for a comparison. We find that the noise resistance of the BarRat method is essentially consistent with that of the MaxEnt method, at least in this test. When $\delta < 10^{-4}$, both methods can accurately recover the three-peak structure of the true solution. When $\delta > 10^{-4}$, both the upper and lower Hubbard bands at $\omega = \pm 2.5$ disappear gradually as $\delta$ increases. One disadvantage of the MaxEnt method compared to the BarRat method is that, when $\delta > 10^{-4}$, the central quasiparticle peak at $\omega = 0$ could split into multiple smaller peaks. If the Prony's approximation is activated for denoising, the BarRat method can yield more distinct upper and lower Hubbard bands when $\delta = 10^{-4}$. However, when $\delta > 10^{-4}$, this combination (BarRat + Prony's approximation) also fails to reproduce the Hubbard bands. At first glance, this test suggests that the Prony's approximation is of little avail in improving the noise resistance of the BarRat method. Is that true? To clarify this question, more tests are highly desired.

\subsection{Denoising}

Does denoising by the Prony's approximation actually help the BarRat method? We need quantitative benchmark results to make the correct judgment. To achieve this goal, we have to define a variable to measure the distance between the calculated and exact (or ideal) spectral functions. That is the error function~\cite{PhysRevB.110.035154}:
\begin{equation}
\label{eq:error}
\text{err}(A) = \int d\omega~\left|A_{\text{calc}}(\omega) - A_{\text{true}}(\omega)\right|.
\end{equation}
We adopt Test T$_{3}$ as an example again. The noise levels are fixed to $\delta$ = $10^{-2}$, $10^{-4}$, and $10^{-6}$, which imply large, intermediate, and small noise, respectively.

At first, we examine the error functions of the BarRat and MaxEnt methods. As shown in Figure~\ref{fig:denoise}, when $\delta = 10^{-2}$ and $10^{-4}$, the error values of the BarRat method (solid lines in the figure) are larger than those of the MaxEnt method (dashed lines in the figure). This indicates that under the same conditions, the solution of the MaxEnt method is closer to the true solution. However, as $\delta$ decreases, the difference in accuracy between the BarRat and MaxEnt methods narrows. As $\delta = 10^{-6}$, the error value of the BarRat method becomes smaller than that of the MaxEnt method, making the BarRat method superior for high-precision Matsubara data.

Next, we consider the denoising effect of the Prony's approximation. As stated above, the precision of the Prony's approximation (i.e., its denoising capability) is controlled by the $\epsilon$ parameter [see Eq.~(\ref{eq:prony})]. On one hand, if $\epsilon$ is small, the interpolated Green's functions after the Prony's approximation will be closer to their original values. On the other hand, if $\epsilon$ is large, the Prony's approximation tends to suppress data fluctuation and produce a smooth curve. For a given noise level $\delta$, we adjust the value of $\epsilon$ and observe how the error value of the BarRat method + Prony's approximation changes with respect to $\epsilon/\delta$. When $\delta = 10^{-2}$ and $10^{-4}$, the Prony's approximation is quite effective. We can see that the error values by the BarRat method + Prony's approximation (filled circles in Fig.~\ref{fig:denoise}) are comparable to those of the MaxEnt method, but the corresponding $\epsilon/\delta$ should not be too large. If $\epsilon/\delta$ is too large, the error values could be significant, possibly even larger than the one by the BarRat method without denoising. The reason is quite simple: a large $\epsilon$ parameter causes the Prony's approximation to overfit. When the noise level is small (such as $\delta = 10^{-6}$), the Prony's approximation has almost no effect. The error values hardly change with $\epsilon/\delta$, practically coinciding with the error value of the BarRat method without denoising. Therefore, we can conclude that denoising by the Prony's approximation is more suitable for applications with high noise levels and should primarily avoid overfitting. If the noise level of the data is low or the data is noise-free, the Prony's approximation is not necessary.

\subsection{Size of input data\label{subsec:size_of_data}}

The computational accuracy of the BarRat method depends on not only the noise level but also the size of input data. Once the input data is insufficient, the accuracy of the BarRat method will deteriorate. To demonstrate this statement, we perform extensive tests with Test T$_{3}$. The noise level of the input data is fixed to $10^{-10}$. The error value is evaluated by Eq.~(\ref{eq:error}). The number of Matsubara data points, $N_w$, changes from 10 to 400. Besides, the other computational parameters are consistent with the previous tests. As is seen in Fig.~(\ref{fig:dataset}), when $10 < N_w < 20$, the error values of the BarRat method are much larger than those of the MaxEnt method. If $N_w$ is further increased to 30, the error values would decrease sharply and be smaller than those of the MaxEnt method. When $N_w \geq 100$, the error values essentially stabilize and approach a constant. Further increasing in $N_w$ won't improve the computational accuracy. Examining other tests yields similar results. Therefore, the optimal size of input data for the BarRat method should be around 100.

\subsection{Efficiency}

\begin{table*}[ht]
\caption{Computational efficiencies for selected analytic continuation methods. All the calculations are done by using the \texttt{ACFlow} toolkit~\cite{Huang:2022}. For each test, we repeat the calculations at least 10 times and calculate the averaged elapsed time. In the analytic continuation of Nambu Green's function, only the normal component is treated. For T$_{3}$, T$_{5}$, and T$_{11}$, the noise levels of input data are fixed to $10^{-6}$. For the Prony's approximation, the $\epsilon$ parameter is set to $10^{-6}$. In the SPX calculations, the number of poles is 2000 for continuous spectra and 10 for discrete spectra. The number of Monte Carlo samplings is fixed to $10^5$, which is enough to obtain accurate solutions. Please see Ref.~[\onlinecite{Huang:2022}] for more details about the SPX method. \label{tab:speed}}
\begin{tabular}{lllllll}
\hline
\hline
    Test & Model                  & MaxEnt & BarRat  & BarRat + Prony's approximation & SPX & Section \\
\hline
    T$_{3}$   & Lorentzian model       & 9.5 s  & 0.013 s & 0.016 s & $> $1 hr  & \ref{sec:lorentz_f}\\
\hline
    T$_{5}$   & Pole model             & 43.8 s & 0.019 s & 0.024 s & $> $5 min & \ref{sec:lorentz_f}\\
\hline
    T$_{11}$  & Optical conductivity   & 2.59 s & 0.012 s & 0.032 s & $> $1 hr  & \ref{sec:current}\\
\hline
Nambu Green's function & Anderson impurity model & 32.6 s & 0.03 s  & 0.12 s  & $> $1 hr  & \ref{sec:nambu_g}\\
\hline
\hline
\end{tabular}
\end{table*}

Aside from noise resistance and accuracy, the computational efficiency of analytic continuation methods is also crucial. In finite-temperature many-body perturbative calculations~\cite{PhysRev.139.A796,Aryasetiawan_1998,RevModPhys.74.601,PhysRevB.91.235114}, quantum Monte Carlo calculations~\cite{gubernatis_kawashima_werner_2016,RevModPhys.73.33}, and non-local dynamical mean-field theory simulations of lattice and condensed matter systems~\cite{RevModPhys.78.865,RevModPhys.77.1027,RevModPhys.90.025003}, a large number of momentum-dependent lattice Green's functions and self-energy functions may require analytic continuations. In this case, high-performance analytic continuation methods become indispensable. Although the SAC method and its variants exhibit excellent accuracy and noise resistance, their efficiency is too low to be extensively used in analytic continuation tasks~\cite{beach2004,PhysRevB.57.10287,PhysRevE.94.063308,PhysRevB.76.035115,PhysRevX.7.041072,SHAO20231}. Currently, only the MaxEnt method can realize a good balance between computational accuracy and efficiency~\cite{JARRELL1996133}.

In previous tests, we already demonstrated that the computational accuracy and applied range of the BarRat method are not inferior to the MaxEnt method. Now we will concentrate on the computational efficiency of the BarRat method. We select four typical examples. They are solved by the MaxEnt method, the BarRat method, the BarRat method + Prony's approximation, and the SPX method~\cite{PhysRevB.108.235143}. The elapsed time is recorded and shown in Table~\ref{tab:speed}. It is evident that the BarRat method takes the least amount of time in all the tests. Furthermore, even if the Prony's approximation is activated for denoising, there is still no significant increase in elapsed time. The MaxEnt method is quite low in efficiency when compared to the BarRat method. In the four tests, the MaxEnt method is at least 100 times slower than the BarRat method. Just as expected, the SPX method takes the longest time. It spends too much time generating millions of random configurations and calculating their contributions to the spectrum.

\subsection{Relations with the other methods}

As mentioned before, the BarRat method was inspired by the PES method~\cite{PhysRevB.107.075151} and the MPR method~\cite{PhysRevB.110.035154,PhysRevB.110.235131}. Below, we will discuss their similarities and differences.

The PES method relies on the pole representation for the Matsubara Green's function~\cite{PhysRevB.107.075151}. In this method, the barycentric rational function approximation~\cite{S0036144502417715,17M1132409} and the AAA algorithm~\cite{AAA,AAA_Lawson} are also applied. But their roles are limited in the second step, i.e., pole estimation. They are used to roughly estimate the positions of the poles in a casual space, providing initial guesses for subsequent semidefinite relaxation calculations. In principle, they can be replaced by the other optimization algorithms. In the BarRat method, we fully leverage the advantages of the barycentric rational function approximation and the AAA algorithm, avoiding complicated non-convex optimization calculations~\cite{PhysRevB.107.075151}. The Matsubara Green's function can be expressed directly as a rational function in the complex plane, instead of a polynomial that is based on the pole representation.

In the MPR method, there are two instances of the Prony's approximation calculations~\cite{PhysRevB.110.035154,PhysRevB.110.235131}. The first calculation is to represent the Matsubara data in the form of Prony's approximation, followed by a holomorphic transform that maps the complex-valued function defined on the imaginary axis to a function on the unit disk. The second calculation is to extract the positions and weights of the poles for the Matsubara Green's function. It is evident that in this method, the Prony's approximation is fundamental. In contrast, in the BarRat method, the Prony's approximation is merely for denoising and is not mandatory. We can replace it with the other denoising algorithms.

The PA method is an early analytic continuation method that uses continued fractions to interpolate Matsubara Green's functions~\cite{PhysRevB.87.245135,PhysRevB.93.075104,Vidberg1977,PhysRevD.96.036002,PhysRevB.61.5147}. It is extremely sensitive to data noise. Therefore it is rarely applied to the analytic continuation of quantum Monte Carlo simulation data~\cite{gubernatis_kawashima_werner_2016,RevModPhys.83.349,RevModPhys.73.33}. Mathematically, the continued fractions used in the Pad\'{e} approximation and the barycentric rational function are both forms of rational functions. They can be converted into each other. Consequently, the two methods are of the same origin, but their numerical stabilities differ greatly. Previous studies have shown that barycentric rational functions combined with the AAA algorithm are currently the most efficient and stable rational function interpolation algorithms~\cite{AAA,AAA_Lawson}.

\section{Concluding remarks\label{sec:con}}

Analytic continuation of Matsubara Green's functions is a critical need in quantum many-body calculations. In this paper, we present the BarRat method, a simple and efficient approach for solving the analytic continuation problems. This method takes advantage of the barycentric rational functions to directly interpolate Matsubara Green's functions. Then the AAA algorithm is utilized to determine the nodes and weights of the barycentric rational functions. Finally, we can establish analytic expressions for the Matsubara Green's functions in the whole complex plane.

The performance of the BarRat method has been systematically explored through a series of toy models and realistic examples. The calculated results suggested that the BarRat method can accurately reproduce major characteristics of the spectral functions, irrespective of continuous or discrete spectra. This method can be used to handle both fermionic and bosonic Green's functions, irrespective of diagonal or non-diagonal components. In addition, special treatment isn't required for spectral functions without positive definiteness~\cite{PhysRevB.92.060509}. This method exhibits prominent noise tolerance and numerical stability. If the noise level of input data is significant, the Prony's approximation can be adopted to suppress the noise. Most of all, the BarRat method outperforms traditional methods in terms of computational speed, making it a competitive alternative to the MaxEnt method.

In conclusion, the BarRat method provides a fast and reliable analytic continuation tool for researchers to extract physical observables from quantum Monte Carlo simulation data with remarkable accuracy. Future work will concentrate on testing the method's applicability to the analytic continuation of frequency-dependent transport coefficients with non-positive spectral weight~\cite{PhysRevB.95.121104}, potentially in combination with the auxiliary Green's function method~\cite{PhysRevB.108.L220503,PhysRevB.90.041110,Tomczak_2007}. In addition, ongoing development aims to enhance the method's performance further and investigate its integration with the other denoising algorithms.

\begin{acknowledgments}
Li Huang was supported by the National Natural Science Foundation of China (Grants No.~12434009 and No.~12274380) and the National Key Research and Development Program of China (Grant No.~2024YFA1408602). Changming Yue was supported by the National Natural Science Foundation of China (Grant No.~1247041908), the Guangdong Provincial Quantum Science Strategic Initiative (Grant No.~SZZX2401001), and the Guangdong Provincial Key Laboratory of Advanced Thermoelectric Materials and Device Physics (Grant No.~2024B1212010001). Changming Yue also acknowledges support from the research startup grant Y01206254 by Southern University of Science and Technology (SUSTech).
\end{acknowledgments}

\bibliography{rfa}

\end{document}